\documentclass[aps,prb,letterpaper,twocolumn, floatfix]{revtex4}
\usepackage{setspace}

\usepackage{color}
\usepackage{braket}
\usepackage{amsmath}
\usepackage{amssymb}
\usepackage[pdftex]{graphicx}

\begin{document}
\title{Signatures of the superfluid to Mott insulator transition in equilibrium and in dynamical ramps}
\author{D. Pekker$^1$, B. Wunsch$^{2,3}$, T. Kitagawa$^3$, E. Manousakis$^4$, A. S. S{\o}rensen$^5$, E. Demler$^3$}
\affiliation{
$^1$Department of Physics, Caltech University, Pasadena, CA 91125, USA\\
$^2$ABB Switzerland Ltd., Corporate Research, Baden CH 5405\\
$^3$Physics Department, Harvard University, Cambridge, MA 02138, USA\\
$^4$Department of Physics, Florida State University, Tallahassee, FL 32306, USA\\
$^5$QUANTOP, Danish Quantum Optics Center and Niels Bohr Institute, DK-2100 Copenhagen \O, Denmark}
\date{\today}
\begin{abstract}
We investigate the equilibrium and dynamical properties of the Bose-Hubbard model and the related particle-hole symmetric spin-1 model in the vicinity of the superfluid to Mott insulator quantum phase transition. We employ the following methods: exact-diagonalization, mean field (Gutzwiller), cluster mean-field, and mean-field plus Gaussian fluctuations. In the first part of the paper we benchmark the four methods by analyzing the equilibrium problem and give numerical estimates for observables such as the density of double occupancies and their correlation function. In the second part, we study parametric ramps from the superfluid to the Mott insulator and map out the crossover from the regime of fast ramps, which is dominated by local physics, to the regime of slow ramps with a characteristic universal power law scaling, which is dominated by long wavelength excitations. We calculate values of several relevant physical observables, characteristic time scales, and an optimal protocol needed for observing universal scaling.
\end{abstract}
\maketitle

\section{Introduction}
Recent experimental progress on trapping, control, and imaging of ultra-cold atoms has advanced to a similar level of accuracy as high fidelity simulation of the Bose-Hubbard model over a wide parameter regime. Most intriguing aspect of experiments reported in Refs.~\onlinecite{Bakr2009, Bakr2010, Sherson2010}, is the ability to read out the state of the atoms with single site resolution. Although the static properties of the Bose-Hubbard model have been studied extensively using various numerical techniques, most significantly Quantum Monte Carlo (QMC), the same is not true regarding its dynamic properties. In particular, we are interested in the time dependent behavior of the Bose-Hubbard model undergoing parametric drive (i.e.~tuning of the hopping matrix element and the onsite interaction in time).

One aspect of parametric drive that has received a lot of theoretical attention is ramping a system across a phase transition~\cite{Dziarmaga2010, Polkovnikov2011}. The classical version of this problem was originally addressed by Kibble and Zurek, who observed that for a thermodynamic phase transition as a system is driven from a disordered phase into an ordered phase, different regions of the system order independently, thus, entrapping topological excitations (e.g. vortices or hedgehogs). Further, the density  of topological excitations that is entrapped depends on how fast and how far into the ordered phase the system is driven~\cite{Kibble1976, Zurek1985}. Later, this analysis was extended to dynamical crossing of a quantum phase transition~\cite{Polkovnikov2005, Zurek2005, Cherng2006}. There, it was observed that dynamics becomes universal if one parametrically drives the system in the vicinity of a quantum critical point. Moreover, the signatures of this universal dynamics will appear as universal power law behavior of observables like the number of collective modes excited and the energy pumped into the system as a function of how rapidly the system is ramped across the phase transition~\cite{Polkovnikov2008, DeGrandi2010, Rams2011, Natu2011}. 

The Bose Hubbard-Model and the corresponding experiments using ultracold bosons in optical lattices, are a natural test bed for studying universal dynamics both theoretically and experimentally. From the experimental perspective, studying bosons in lattices is attractive for several reasons. First, it is possible to prepare the atoms in very ``cold" superfluid state, thus, allowing one to concentrate on studying the quantum dynamics. Second, the timescales available in experiments are favorable for observing dynamics: the extrinsic timescale for atom loss is long compared to the intrinsic Bose-Hubbard timescales which are themselves long compared to the time needed to tune the system parameters and make observations. Finally, using the ultracold gas microscope~\cite{Bakr2009, Sherson2010}, it is possible to see the occupation number (or at least its parity) of individual lattice sites, thus, obtaining a very sensitive probe of the dynamics. The parity probe can be thought of as counting defects, i.e. sites with too many or too few bosons as compared to the average occupation number. From the theoretical perspective, the equilibrium properties of the Superfluid-Mott transition are well understood. However, there is a lack of tools that can be used for studying the dynamics of interacting systems. The Bose-Hubbard model, thus, provides theory with a challenge, as well as a hope of future comparison with experiments.  

In this manuscript, we shall apply a number of numerical approaches to investigate the properties of the superfluid to Mott insulator transition in two dimensions, at commensurate filling, both in and out of equilibrium. In particular, we shall begin by investigating the zero temperature equilibrium properties such as density of defects and their correlation functions. Next, we shall look at the same properties in dynamic ramps across the phase transition starting from the zero temperature equilibrium state. We shall use (1)  Exact Diagonalization (ED), (2) Mean-Field (MF) theory, (3) Cluster-Mean-Field (CMF), and (4) Mean-Field theory with Gaussian fluctuations (MF+G) methods. The MF+G method is an extension of the normal mode analysis of Refs.~\cite{Altman2002,Huber2007} to include time dependence of the mean field on top of which the normal modes are constructed. We shall apply these methods to the Bose-Hubbard model, as well as the closely related spin-1 quantum rotor model. Our goals are two-fold (a) to model the current generation of experiments, which take place on relatively short timescales and (b) to comment on what is needed to observe universal scaling in dynamics experiments at correspondingly longer timescales. We shall take the strategy of first making the confidence building measure of comparing the results of these methods with QMC in equilibrium at zero temperature. In building up our confidence, we obtain relatively simple means for computing experimentally measured quantities including the defect densities and defect correlation functions. Next, we shall employ these methods to study the dynamics during parametric ramps to the Mott insulating phase starting with equilibrium superfluid at zero temperature. 

In order to study dynamics numerically, it is necessary to make some approximations. Indeed, our first approximation will be to simplify the Hamiltonian. We do this in two ways. First, we truncate the Hilbert space to three states per site. That is, given the average density of $n_0$ bosons per site, we truncate the Hilbert space on each site to the three states corresponding to occupation by $\{n_0-1,n_0, n_0+1\}$ bosons. Second, for the case of MF+G method, we concentrate on the case of large $n_0$, which reduces the Hubbard model to the quantum rotor model and introduces an exact particle-hole symmetry. Making these two approximations does not effect the universality class of the phase transition, and, thus, should preserve the universal dynamics. Moreover, as we shall show these approximations do not change the non-universal properties qualitatively. Finally we remark that we will focus on homogeneous systems, thus, avoiding the question of the redistributions of atoms (and energies) in the trap. 

Our main results are as follows. For understanding equilibrium properties, the methods we consider all have strength and weaknesses. All of the methods show qualitative agreement for computing non-universal properties. The particular tests we looked at were (1) finding the phase boundary between superfluid and Mott insulator, (2) computing the defect density, and (3) defect -defect, particle-hole, and particle-particle correlation functions. The location of the phase boundary can be computed using MF and CMF methods. We find that both of these methods obtain a similar phase boundary, however the CMF method with large clusters provides a significant improvement over the MF method when compared to the exact Quantum Monte Carlo predictions. Likewise, all of the methods do a reasonable job in calculating the number of defects in equilibrium as a function of the tuning parameter. However, they show quantitative difference amongst themselves. These differences shrink as the cluster size (for CMF method) and system size (for ED method) increase. Finally, we can also apply these methods (CMF, ED, MF+G) for calculating short range equilibrium correlation functions. Here, we again find quantitative differences, but qualitatively similar results between the different methods. Only the MF+G method can be used for computing long range correlations functions, and we use it to find the diverging length scale at the transition and give some estimates on the quantitative values of the $g_2$ function near the transition (which could be used for comparison with experiment).

For understanding dynamics, we find that there are two regimes: fast and slow. To define the timescales, we first define the energy scales in the problem: $J$ is the hopping matrix element and $U$ is the on-site interaction matrix element (see Section~\ref{sec:model} for the exact definition of the models that we study).  For fast ramps, or short timescales as compared to $2 \pi \hbar/J$ (we operate near the phase transition where $2 \pi \hbar/J \sim 4 z(2\pi\hbar/U)$), short-range physics dominants and all methods produce qualitatively similar features. In particular we see a strong response on timescales of $(2\pi\hbar/U)$, followed by prominent oscillations with period $\sim 2(2\pi\hbar/U)$. This short timescale is associated with the on-site repulsion energy scale, and appears naturally from the perspective of collective modes as a strong peak in the density of states of both the phase and the amplitude modes that occurs near the lines $k_x\pm k_y=\pi$. For slow ramps or long timescales, long wavelength modes become important, although the corresponding density of states is much smaller. We note that what we call the fast regime was studied within the MF method in Ref.~\cite{Natu2011}. The only method that can capture non-local entanglement and correlations that we have is the MF+G method, which shows significant deviations from the other methods. Within the MF+G method we find a crossover to the universal power-law scaling regime which is not seen by MF, CMF, nor ED methods. We find that for ramps that start in the superfluid and end deep in the Mott insulator, the crossover to universal scaling occurs for ramps of $\sim 15*(2\pi \hbar/J_c)$ where $J_c$ is the hopping matrix element at the phase transition. For the experimental conditions of Ref.~\cite{Bakr2010}, $^{87}$Rb in 1360~nm lattice, the corresponding timescale is $\sim 1\, \text{s}$. It is possible to achieve crossover to the universal scaling regimes for faster ramps by starting at the QCP, but experimentally that is detrimental as it is difficult to prepare the system at the QCP. Finally, we show that by ending the ramp deep in the Mott insulator, the excitations created near the QCP are converted into defects which can be detected experimentally.

The manuscript is organized as follows. In Section~\ref{sec:model} we define the Bose-Hubbard and the quantum rotor models and truncate the Hilbert space. We introduce the four methods: ED, MF, CMF, and MF+G in Section~\ref{sec:method}. We study the equilibrium properties of the Bose-Hubbard and quantum rotor models using these methods in Section~\ref{sec:equilibrium}, and dynamics properties in Section~\ref{sec:dynamics}. Finally, we draw conclusions and discuss implications of our results for future experiments in Section~\ref{sec:discussion}. The main body of the Paper is supplemented by three appendices in which we describe the details of the equilibrium and dynamics of the MF+G model and the connection to experiment.

\begin{figure}
\begin{minipage}[b]{0.3cm}
	{\bf (a)}
	
	\vspace{4.5cm}
\end{minipage}
\begin{minipage}[t]{8.2cm}
	\begin{flushleft}
		\includegraphics[scale=0.8]{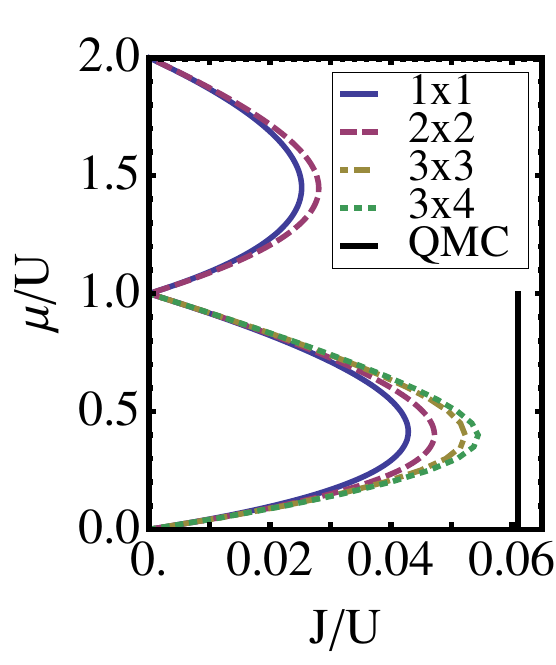} \\
	\end{flushleft}
\end{minipage}\\ \vspace{0.5cm}
\begin{minipage}[b]{0.3cm}
	{\bf (b)}
	
	\vspace{4.cm}
\end{minipage}
\begin{minipage}[t]{8.2cm}
	\begin{flushleft}
		\includegraphics[scale=0.8]{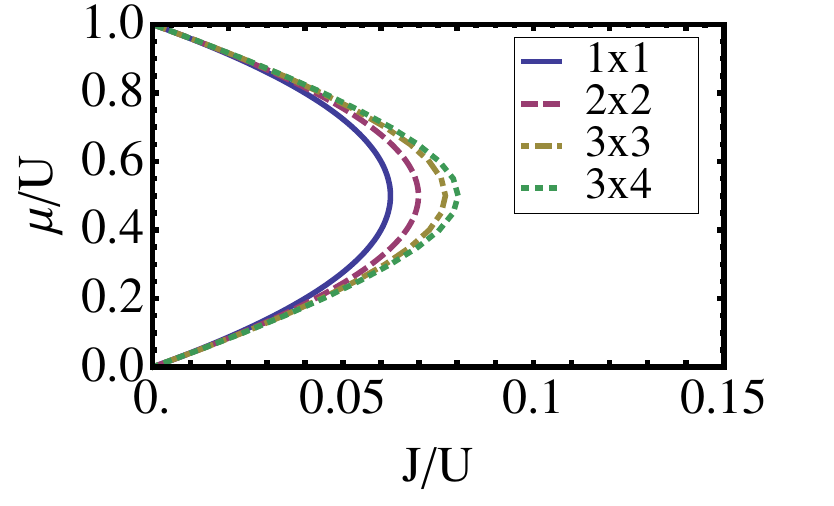}
	\end{flushleft}
\end{minipage}
\caption{Comparison of Mott insulator --- Superfluid phase boundary for the (a) Bose-Hubbard model and the (b) spin-1 quantum rotor model, in the chemical potential-hopping matrix element plane, obtained using mean field (cluster size 1x1) and cluster mean field methods (cluster size ranging from 2x2 to 3x4) in 2D. The vertical black line in (a) indicates the location of the tip of the $n=1$ lobe obtained from quantum Monte Carlo (QMC)~\cite{Capogrosso-Sansone2007}. In order to capture the $N=2$ Mott lobe, we extended the basis for the 1x1 and 2x2 calculations to include 3 particles per site. Quantitatively the extent of the tip of the  $n=1$ lobe for the Bose-Hubbard model is: $J_H/U_H=\{0.043,0.047,0.052,0.054,0.061\}$ for the \{1x1 MF, 2x2 CMF, 3x3 CMF, 3x4 CMF, QMC\} methods, respectively. For the spin-1 quantum rotor model the tip extent is: $J/U=\{0.0625, 0.070, 0.077, 0.080\}$ for the \{1x1 MF, 2x2 CMF, 3x3 CMF, 3x4 CMF\} methods, respectively.}
\label{fig:PhaseBoundary}
\end{figure}

\section{Models}
\label{sec:model}
The Bose-Hubbard Hamiltonian is
\begin{align}
H=-J_H \sum_{\langle i j \rangle} a_i^\dagger a_j + \frac{U_H}{2} \sum_{i} n_i (n_i-1) - \sum_i \mu_i n_i,
\label{eq:HBH}
\end{align}
where $a_i^\dagger$ and $a_i$ are the boson creation and annihilation operators at site $i$, $n_i=a_i^\dagger a_i$ is the boson number operator, and $\mu_i$ is the site dependent chemical potential that describes the trap. In the homogeneous setting, this model supports two types of ground states: Superfluid and Mott insulator~\cite{Sachdev1999}.  The phase diagram in the $\mu/U_H$ --- $J_H/U_H$ plane consists of a number of Mott Insulating lobes surrounded by a Superfluid~\cite{Fisher1989} (see Fig.~\ref{fig:PhaseBoundary}). As we are ultimately interested in ramping between superfluid and Mott phases, we will mostly consider systems with integer average density (however, we shall also comment on the effects of the trap). 

Although the properties of the Bose-Hubbard model are the ultimate aim of this paper, it is difficult to obtain any analytical results regarding dynamics for this model in its original form. Therefore, in order to compare various methods, we shall use both the Bose-Hubbard model and its simplified cousin the quantum-rotor like model. The quantum rotor model is described by the Hamiltonian 
\begin{align}
H=- J \sum_{\langle i, j \rangle} S^+_i S^-_j + \frac{U}{2}\sum_i \left(S^z_i\right)^2 - \sum_i \mu_i S^z_i,
\label{eq:HSpin1}
\end{align}
and is closely related to the Bose-Hubbard model. In both cases, we shall truncate Hilbert space to three states per site (with the exception of the calculation of the Bose-Hubbard model phase diagram, where we extend the basis in order to capture the $N=2$ Mott lobe in addition to the $N=1$ Mott lobe). That is, for the quantum rotor model the site basis will corresponds to the eigenstate of the $S^z_i$ operator $\{|-1\rangle_i,|0\rangle_i,|1\rangle_i\}$. Similarly, for the Bose-Hubbard case, we use the eigenstates of the number operator $\{|n_0-1\rangle_i,|n_0\rangle_i,|n_0+1\rangle_i\}$ as the site-basis.  The main ingredient that differentiates the Hubbard and the rotor models is that the later is symmetric between $|n_0-1\rangle_i$ and $|n_0+1\rangle_i$. However, we note that in the limit of large $n_0$ the two models become identical under the identification $J = n_0 J_H$ and $U = U_H$. We shall ignore particle-hole asymmetric terms on the basis that they will not effect the properties of the phase transition. Furthermore, the asymmetric terms will be absent at large fillings. Regardless of the issues of the Hilbert space truncation and asymmetries, the low energy, long wavelength theory which is essential for understanding the scaling in the vicinity of the Superfluid-Mott Insulator transition is identical in these two models~\cite{Sachdev1999}.

\section{Methods in and out of equilibrium}
\label{sec:method}
The goal of this section is to establish the various approximate methods that can be used for treating dynamics. In particular, we investigate the use of exact diagonalization (ED) on small lattices, Weiss-type mean field theory on single sites (MF) as well as on small clusters (CMF), and mean field theory augmented by Gaussian fluctuations (MF+G).

\subsection{Exact Diagonalization (ED)}
Exact diagonalization is perhaps the most straightforward of the methods available for exploring dynamics. The main disadvantage of ED is the exponential increase of its computational complexity as system size is increased. 

We diagonalize both the quantum rotor model and the Bose Hubbard model on small lattices with periodic boundary conditions. For the case of the Bose-Hubbard model the total particle number commutes with the Hamiltonian Eq.~\eqref{eq:HBH}, and therefore we work at a fixed total particle number. Similarly, for the case of the rotor model, the conserved quantity that we fix is the total $S^z$. An additional consideration is the truncation of the site-basis. Although, for the Hubbard model, at fixed total particle number, the Hilbert space is already finite, we find that reducing the size of the Hilbert space further by truncating the site-basis to three elements does not strongly influence the observables such as the defect density [as we demonstrate in the next Section]. On the other hand, for the case of the quantum rotor model, fixing the total $S^z$ does not fix the size of the basis. However, we again find that truncating the site basis to three states does not significantly alter the computed observables as compared to including more site-basis elements.

Using ED, it is possible to look at both static and dynamic properties. The static properties are obtained by finding the eigenstate with the smallest eigenvalue of the Hamiltonian in the reduced Hilbert space
\begin{align}
H_\text{reduced}=\sum_{lm} |\phi_l\rangle \langle \phi_l | H | \phi_m \rangle \langle \phi_m|,
\label{eq:HBHreduced}
\end{align}
where $|\phi_l\rangle$ are the basis vectors that make up the reduced Hilbert space. Dynamics are obtained by solving the Schrodinger evolution equation for the wave function in the reduced Hilbert space
\begin{align}
i \hbar \, \partial_t  | \psi_\text{reduced}(t) \rangle = H_\text{reduced}(t) |\psi_\text{reduced}(t) \rangle.
\end{align}
As initial condition, $|\psi_\text{reduced}(t=0) \rangle$,  we use the eigenstate with the lowest eigenvalue obtained for the initial Hamiltonian $H_\text{reduced} (t=0)$.

The properties that can be studied using ED are local observables, as the range of correlation functions is limited by system size. In particular we will look at the defect density and next-nearest neighbor correlations (i.e. particle-particle, particle-hole, hole-hole), see Table~\ref{table:Measure} for an explicit list.  It is important to note that ED cannot shed light on long range correlations, and therefore phase boundaries. 

\begin{table*}
\begin{tabular}{c|c|c}
Model & Spin-1 & Bose Hubbard \\
\hline
defect density $P_d$ & $(S^z_i)^2$ & 
$
\begin{array}{cc} 
\frac{1}{2}\left(1+\text{Pr}_i\right) & n_0 \in \text{odd} \\
\frac{1}{2}\left(1-\text{Pr}_i\right) & n_0 \in \text{even} 
\end{array} $ \\[15pt]
\hline
defect correlation $g_2(l)$ & $(S^z_i)^2 (S^z_{i+l})^2$ & 
$\begin{array}{cc} 
\frac{1}{4}\left(1+\text{Pr}_i\right) \left(1+\text{Pr}_{i+l}\right) & n_0 \in \text{odd} \\
\frac{1}{4}\left(1-\text{Pr}_i\right) \left(1-\text{Pr}_{i+l}\right) & n_0 \in \text{even} 
\end{array} $ \\[15pt]
\hline
particle-particle correlation $g_{p-p,2}(l)$ & $\frac{1}{4} \, S^z_i (1+S^z_i)\, S^z_{i+l} (1+S^z_{i+l})$ & 
$n_{p,i} n_{p,i+l}$ \\[5pt]
\hline
particle-hole correlation $g_{p-h,2}(l)$ &  $\frac{1}{4} \, S^z_i (1+S^z_i)\, S^z_{i+l} (1-S^z_{i+l})$  &
$n_{p,i} n_{h,i+l}$ \\[5pt]
\hline
hole-hole correlation $g_{h-h,2}(l)$ &  $\frac{1}{4} \, S^z_i (1-S^z_i)\, S^z_{i+l} (1-S^z_{i+l})$  &
$n_{h,i} n_{h,i+l}$\\[5pt]
\hline
\end{tabular}
\caption{Explicit list of operators that correspond to observables that we compute in the Spin-1 and the Bose-Hubbard models. $\text{Pr}$ is the parity operator and is $+1$ if the number of bosons on site $i$ is even, and $-1$ if it is odd and $n_0$ is the average filling; $n_{p,i}=\text{max}(a^\dagger_i a_i-n_0,0)$ is the particle number operator that counts the excess in the number of bosons on site relative to $n_0$, while $n_{h,i}=\text{max}(n_0-a^\dagger_i a_i,0)$ is the hole number operator that counts the deficit in the number of bosons.}
\label{table:Measure}
\end{table*}

\subsection{Mean field (MF) and Cluster mean field (CMF)}
We use both MF and CMF methods to investigate statics and dynamics of both models. The key advantage of the mean field theories over ED is that they take some long range correlations into account in the form of the order parameter. Thus CMF forms a complimentary approach for probing all the local observables that are available to ED, but in addition can be used to locate phase boundaries and probe the dynamics of the order parameter. As the cluster size increases, CMF should become asymptotically exact, although this regime is difficult to reach in practice.

To apply the CMF method, we begin by partitioning the system into a set of clusters, e.g. we split up the grid of $m x \times  n y$ sites into $x y$ clusters of size $m\times n$. We note that MF can be thought of as a special case of CMF, with a $1\times1$ cluster. The cluster mean field Hamiltonian is similar to the ED Hamiltonian, but with two important differences. First, the particle number ($S^z$ for the quantum rotor case) inside the cluster is not conserved, so we must include basis vectors with different particle numbers in our truncated Hilbert space. Second, instead of periodic boundary conditions we couple the external sites of the cluster to the mean field on neighboring clusters, thus making the MF/CMF solutions consistent. For the uniform case, the cluster being diagonalized must be consistent with itself, thus making the theory self-consistent.

For the case of the quantum rotor model, the coupling to the mean field is obtained by supplementing the Hamiltonian of Eq.~\eqref{eq:HSpin1} with the boundary condition
\begin{align}
-t \sum_{i \in \partial, j \in x} S_i^{+} \langle S_j^{-}\rangle + h.c.,
\end{align}
where $i$ runs over all sites at the boundary of the cluster being diagonalized and $j$ runs over all sites that are nearest neighbors of $i$ but are external to the clusters being diagonalized. We remark that for the uniform case, in which the cluster must be consistent with itself, we must be careful in counting the number of $j$-sites (e.g. corner sites still have two neighbors in 2D). Similarly, for the Bose-Hubbard case, we supplement the Hamiltonian over the truncated Hilbert space Eq.~\eqref{eq:HBH} with
\begin{align}
-t \sum_{i \in \partial, j \in x} b_i^{\dagger} \langle b_j\rangle + h.c.\,.
\end{align}

To study dynamics, we evolve the wave functions and the order parameters in a consistent way. Explicitly, consider a pair of neighboring clusters $A$ and $B$. Each time step can be broken down into two parts. In the first part, we advance the wave function on each cluster in time using the corresponding Schrodinger equation of motion
\begin{align}
|\psi_A(t+\delta)\rangle&= |\psi_A(t)\rangle + \frac{\delta}{i \hbar} H_A(\Psi_B(t), t) |\psi_A(t)\rangle,\label{eq:S1}\\
|\psi_B(t+\delta)\rangle&= |\psi_B(t)\rangle + \frac{\delta}{i \hbar} H_B(\Psi_A(t), t) |\psi_B(t)\rangle,
\label{eq:S2}
\end{align}
where the Hamiltonian for cluster $A$ depends on the order parameter in cluster $B$ and vice versa.
In the second part, we recompute the order parameters (corresponding to the operator $\hat{\Psi}$) in each cluster using the new wave functions  
\begin{align}
\Psi_A(t+\delta) = \langle \psi_A(t+\delta) | \hat{\Psi} | \psi_A(t+\delta)\rangle,\label{eq:S3}\\
\Psi_B(t+\delta) = \langle \psi_B(t+\delta) | \hat{\Psi} | \psi_B(t+\delta)\rangle.
\label{eq:S4}
\end{align}

Before proceeding to the mean field plus Gaussian fluctuations (MF+G), we develop a useful parametrization for the uniform MF solution of the quantum rotor model with the site Hilbert space truncated to three state $|-\rangle$, $|0\rangle$, $|+\rangle$. We begin by pointing out that the MF solution corresponds to the product wave function
\begin{widetext}
\begin{align}
|\psi(\theta, \phi) \rangle&=\prod_i |\psi(\theta, \phi) \rangle_i=\prod_i
\Big[ e^{i\phi} \cos\left(\frac{\theta}{2}\right) |0\rangle_i 
+\frac{e^{-i\phi}}{\sqrt{2}} \sin\left(\frac{\theta}{2}\right)
\left(|+1\rangle_i+|-1\rangle_i\right)\Big],
\end{align}
\end{widetext}
where $\theta$ and $\phi$ are variational parameters that determine the wave function, and $\cos(2\phi) \sin(\theta)/\sqrt{2}$ corresponds to the superfluid order parameter. We have specifically not included any parameters to tune the weight of the $|+\rangle$ state with respect to the $|-\rangle$ state as we shall use this parametrization exclusively at the particle-hole symmetric point ($\mu=0$ in Eq.~\eqref{eq:HSpin1}). The equations of motion Eqs.~\eqref{eq:S1}-\eqref{eq:S4} correspond to the extremum of the effective Lagrangian $L_\text{eff}=\langle \psi(\theta, \phi) | i \partial_t - H | \psi(\theta, \phi) \rangle$ with respect to the $\theta(t)$ and $\phi(t)$. Carrying out the extremization procedure, we obtain the equations of motion
\begin{align}
\dot{\phi}=\frac{U}{4}-J z \cos^2(2\phi)\cos(\theta),\label{eq:MFT1}\\
\dot{\theta}=-2 J z \sin(\theta) \sin(2\phi) \cos(2\phi),
\label{eq:MFT2}
\end{align}
where $z$ is the coordination number (e.g. for 2D square lattice $z=4$). The stationary solutions of these equations of motion correspond to the ground state configurations. The superfluid ground state occurs for $4 J z > U$ while the Mott Insulator occurs in the complimentary regime $4 J z < U$, and the corresponding solutions are
\begin{align}
\theta&=\cos^{-1}\left(\frac{U}{4 z J}\right);  \quad \phi=0 & \text{for } U<4zJ,\\
\theta&=0;  \quad \dot{\phi}=U/4-J z \cos^2(2\phi) & \text{for } U\geq4zJ.
\label{eq:MFgs}
\end{align}
Although the evolution of $\phi$ is seemingly innocuous for the mean-field Mott insulating ground state, we shall revisit it when obtaining the Gaussian fluctuations on top of the Mott insulator.

\subsection{Mean field with Gaussian fluctuations (MF+G)}
Having obtained the mean field, we look at Gaussian excitations on top of it. Here, we follow Refs.~\cite{Altman2002, Huber2007} to setup the spin-wave theory of the truncated quantum rotor model. We focus on this model, as it is significantly more straight forward than the Bose-Hubbard model due to the absence of the bosonic factors. There is some additional simplification due to working at $\langle S^z\rangle=0$ (equivalent of one particle per site average filling), which results in an explicit particle-hole symmetry. 

We begin by introducing the Schwinger-boson-like creation operators  $\{t_{-,i}^\dagger,t_{0,i}^\dagger, t_{+,i}^\dagger\}$ which create the states $|-1\rangle_i$, $|0\rangle_i$, $|+1\rangle_i$ when acting on an ``empty" site. A more convenient  representation can be obtained by rotating these operators via
\begin{widetext}
\begin{align}
 \left(\begin{array}{c} b_{0,i}\\  b_{\alpha,i}\\b_{\phi,i} \end{array}\right) &=M\left(\begin{array}{c} t_{0,i}\\  t_{+,i}\\t_{-,i} \end{array}\right); &
 M&=e^{i \chi}\left(\begin{array}{c c c }
  \cos(\frac{\theta}{2}) e^{i \phi}& \frac{1}{\sqrt{2}} \sin(\frac{\theta}{2}) e^{-i \phi}    & \frac{1}{\sqrt{2}} \sin(\frac{\theta}{2}) e^{-i \phi}\\  
   \sin(\frac{\theta}{2}) e^{i \phi} &  -\frac{1}{\sqrt{2}} \cos(\frac{\theta}{2}) e^{-i \phi} &  -\frac{1}{\sqrt{2}} \cos(\frac{\theta}{2}) e^{-i \phi}\\
   0 & \frac{1}{\sqrt{2}} e^{-i \phi}& -\frac{1}{\sqrt{2}} e^{-i \phi}
  \end{array}\right),
  \label{eq:M}
\end{align}
\end{widetext}
so that $b_0^\dagger$ corresponds to creating the ``mean field" state, while $b_\alpha^\dagger$ and $b_\phi^\dagger$ correspond to creating the two states orthogonal to the mean field state. The labels $\alpha$ and $\phi$ have been chosen to indicate amplitude and phase modes. In this coordinate transformation matrix $M$, we have left the phases $\phi$ and $\chi$ as free parameters, although in equilibrium the phases $\phi$ and $\chi$ can be set to zero. We include them here in order to provide enough flexibility for the dynamical solutions. 

The operator $b_\alpha^\dagger$ corresponds to an amplitude excitation, while $b_\phi^\dagger$ corresponds to a phase excitation, a fact that we can demonstrate by considering the change in the order parameter if we perturb the mean-field $b_0^\dagger$ in the $b_\alpha^\dagger$ or $b_\phi^\dagger$ direction.
\begin{align}
&\langle b_0 + \epsilon^* b_\alpha | S^+ | b_0^\dagger + \epsilon b_\alpha^\dagger \rangle \nonumber \\
&\quad\approx \frac{1}{\sqrt{2}}\sin(\theta) - \sqrt{2}  \cos(\theta) \text{Re}(\epsilon) + O(\epsilon^2)\\
&\langle b_0 + \epsilon^* b_\phi | S^+ | b_0^\dagger + \epsilon b_\phi^\dagger \rangle \nonumber \\
&\quad \approx \frac{1}{\sqrt{2}}\sin(\theta) - i \sqrt{2}  \cos(\theta/2) \text{Im}(\epsilon).
\end{align}
Without applying the perturbations, the order parameter in equilibrium has the value $\frac{1}{\sqrt{2}}\sin(\theta)$. Upon applying a perturbation in the $b_\alpha^\dagger$ direction, we see that the magnitude of the order parameter changes, while a perturbation in the $b_\phi^\dagger$ induces a change of the order parameter phase. Thus we identifying the nature of these operators. As we shall see, the amplitude modes will be built up entirely from $b_\alpha$ and $b_\alpha^\dagger$'s while the phase modes from $b_\phi$ and $b_\phi^\dagger$'s. This separation into amplitude and phase sectors, that do not mix, is a feature of the quantum rotor model, and is not present in the Bose-Hubbard model (where the separation only occurs at small momenta, and at high momenta the modes become mixed). 

Having defined the basis, our next goal is to obtain the effective Hamiltonian up to second order in fluctuations $b_\alpha$ and $b_\phi$. Using the effective Hamiltonian, we can find the ground state wave function, as well as its time evolution. 

Before proceeding to obtain the effective Hamiltonian, we make the approximation that we shall avoid implementing feedback of fluctuations of the amplitude and phase modes back onto the mean field in both statics and dynamics. That is, we shall obtain the ``mean field" $b_0$, described by $\theta(t)$ and $\phi(t)$ from the solution of Eq.~\eqref{eq:MFT1} and \eqref{eq:MFT2}. For the case of static Hamiltonian $\theta(t)$ and $\phi(t)$ will remain fixed, while for the case of the parametrically tuned Hamiltonian they will evolve as a function of time. On top of the mean-field solution, we shall construct the evolution of the fields $b_\alpha$ and $b_\phi$ without any feedback to $\theta(t)$ and $\phi(t)$. Our approximation of not implementing feedback relies on the assumption that fluctuations will only have a small effect on the ``mean field." We comment that implementing feedback must be done correctly using additional machinery such as the Popov approximation in order to ensures that the phase mode remains gapless in the superfluid phase.

Our goal is to expand $H_\text{eff}$ to second order in $b_\alpha$ and $b_\phi$ operators. To do this we convert from the Schwinger boson like formalism to the Holstein-Primakoff like formalism via the replacement $b^\dagger_{0,j} b_{0,j} \rightarrow 1-b^\dagger_{\alpha,j}b_{\alpha,j}-b^\dagger_{\phi,j}b_{\phi,j}$ see Ref.~\cite{Altman2002,Huber2007}. We continue by obtaining each term of the Hamiltonian Eq.~\eqref{eq:HSpin1}, starting with the 
$\frac{U}{2}$ term:
\begin{align}
&\frac{U}{2}\sum_i\left(S^z_j\right)^2 = 
\frac{U}{2}\sum_j \Bigg(\frac{1}{2} \left(1-\cos\left(\theta\right)\right) \nonumber \\
& \quad \quad + \cos\left(\theta\right) b^\dagger_{\alpha,j} b_{\alpha,j}
  +\frac{1}{2} \left(1+\cos\left(\theta\right)\right) b^\dagger_{\phi,j} b_{\phi,j}\Bigg).
\end{align}
Next, we move on to the $J$ term, which first requires us to compute the operators $S^\pm_j$
\begin{align}
\sqrt{2} S^+_j&=\sqrt{2}\left(t^\dagger_{0,j} t_{-,j}+t^\dagger_{+,j} t_{0,j}\right) \\
&=\cos(2\phi) \sin(\theta) b^\dagger_{0,j} b_{0,j}-\cos(2\phi) \sin(\theta) b^\dagger_{\alpha,j}b_{\alpha,j} \nonumber \\
&\quad + i \sin(2\phi) \left(b^\dagger_{\alpha,j} b_{0,j}-b^\dagger_{0,j} b_{\alpha,j}\right) \nonumber \\
&\quad -  \cos(2 \phi) \cos(\theta)\left(b^\dagger_{\alpha,j} b_{0,j}+b^\dagger_{0,j} b_{\alpha,j}\right) \nonumber \\
&\quad+\sin\left(\frac{\theta}{2}\right)\left[e^{-2 i \phi} b^\dagger_{\phi,j}b_{\alpha,j}-e^{2 i \phi} b^\dagger_{\alpha,j}b_{\phi,j}\right] \nonumber \\
&\quad+\cos\left(\frac{\theta}{2}\right)\left[e^{-2 i \phi} b^\dagger_{\phi,j}b_{0,j}-e^{2 i \phi} b^\dagger_{0,j}b_{\phi,j}\right]\\
S^-_j&=\left(S^+_j\right)^\dagger.
\end{align}
In the Hamiltonian the spin raising and lowering operators always appear in the symmetric form 
\begin{align}
S^+_i S^-_j+S^-_iS^+_j = \cos^2(2\phi) \sin^2(\theta) + {\cal S}_\alpha + {\cal S}_\phi,
\end{align}
where 
\begin{align}
{\cal S}_\alpha &=\left[\left(\cos(\theta) \cos(2\phi)+i \sin(2\phi)\right)^2 b^\dagger_0 b^\dagger_0 b_{\alpha,i} b_{\alpha,j} + h.c.\right]\nonumber  \\
&+\left[\frac{1}{4}\left(3+\cos(2\theta)-2 \cos(4\phi) \sin^2(\theta)\right) b^\dagger_{\alpha,j} b_{\alpha,i} + h.c.\right]\nonumber  \\
&-2\cos^2(2\phi) \sin^2(\theta)\left[ b^\dagger_{\alpha,i} b_{\alpha,i} + b^\dagger_{\alpha,j} b_{\alpha,j}\right], \label{eq:Salpha}\\
{\cal S}_\phi &=\left[-\frac{1}{2} e^{4 i \phi} \left(1+\cos(\theta)\right) b^\dagger_0 b^\dagger_0 b_{\phi,i} b_{\phi,j} + h.c.\right]\nonumber\\
&+\left[\frac{1}{2} \left(1+\cos(\theta)\right) b^\dagger_{\phi,j} b_{\phi,i} + h.c.\right]\nonumber\\
&-\cos^2(2\phi)\sin^2(\theta)\left[b^\dagger_{\phi,i} b_{\phi,i} + b^\dagger_{\phi,j} b_{\phi,j}\right].
\label{eq:Sphi}
\end{align}
Here, we have explicitly left terms like $b^\dagger_0 b^\dagger_0$ in the expression for the reason that although we know the amplitude of these terms (which is unity at this level of approximation) we do not know their phase. We will shortly show that the phase associated with these terms may be removed by a specific choice of $\chi$ in transformation Eq.~\eqref{eq:M}, which allows us to make the substitution $b^\dagger_0 b^\dagger_0\rightarrow 1$. We note that in obtaining Eqs.~(\ref{eq:Salpha},\ref{eq:Sphi}) we have dropped terms that are first order in $b_\alpha$ and $b_\phi$ operators (and third order in $b_{0,k}$ operators). The reason for doing this is that these terms involve $b_{0,k}$ operators at finite momentum $k$, which vanish since we assume a spatially uniform mean field. On the other hand, when we consider dynamics of the spatially uniform ($k=0$) mode, the terms that are first order in $b_\alpha$ and $b_\phi$ operators become important. Below Eq.~\eqref{eq:EOMBF}, we show that taking these terms into account we recover the mean field equations of motion Eqs.~\eqref{eq:MFT1} and \eqref{eq:MFT2}. 

\begin{figure*}
\includegraphics[width=\textwidth]{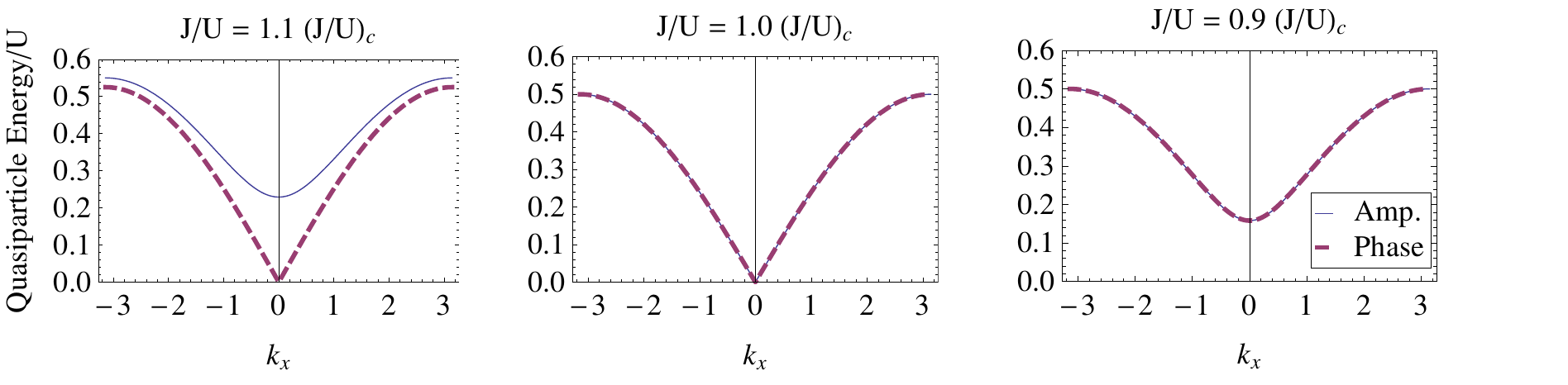}
\caption{Equilibrium dispersion relations~\eqref{eq:dispersion} of the phase (dashed line) and amplitude (solid line) modes at three points in the phase diagram (from left to right: superfluid, QCP, and Mott insulator). The position along the $(J/U)$ axis in the phase diagram is measured relative to the ratio $(J/U)_c$ at the quantum critical point. In plotting these dispersions we set $k_y=0$ and varied $k_x$.}
\label{fig:dispersion}
\end{figure*}

Having obtained all parts of the Hamiltonian in real space, we put them all together and Fourier transform to obtain the complete Hamiltonian up to quadratic order
\begin{align}
H_{\text{eff}}&=\frac{U}{4}\left(1-\cos(\theta)\right) -\frac{J z}{2} \cos^2(2\phi) \sin^2(\theta)\nonumber\\
&\quad\quad \quad\quad \quad\quad \quad\quad\quad\quad \quad\quad +H_{\alpha 2}+H_{\phi 2}.
\label{eq:Heff2}
\end{align}
Here, 
\begin{align}
H_{\sigma 2}&=\frac{1}{2}\sum_k  
\left(\begin{array}{c c} b_{\sigma,k}^\dagger & b_{\sigma,-k} \end{array}\right) 
\left(\begin{array}{c c} 
\alpha_{\sigma,k} & \beta_{\sigma,k} \\
\beta^*_{\sigma,k} & \alpha_{\sigma,k} 
\end{array}\right) 
\left(\begin{array}{c} b_{\sigma,k} \\ b_{\sigma,-k}^\dagger \end{array}\right),
\label{eq:Hsigma2}
\end{align}
are the quadratic Hamiltonians that describe the amplitude ($\sigma=\alpha$) and the phase ($\sigma=\phi$) modes. The coefficients $\alpha_{\sigma,k}$ and $\beta_{\sigma,k}$ are explicitly given by
\begin{align}
\alpha_{\alpha,k} &=\frac{U}{2} \cos(\theta)-2 J z \cos^2(2\phi) \sin^2(\theta) \nonumber \\
&\quad\quad+\frac{\epsilon_k}{4}\left[3+\cos(2\theta)-2 \cos(4\phi)\sin^2(\theta)\right]
\label{eq:alpha:alpha}\\
\beta_{\alpha,k} &=\epsilon_k \left[\cos(\theta)\cos(2\phi)-i \sin(2\phi)\right]^2 e^{2 i \phi}
\label{eq:beta:alpha}\\
\alpha_{\phi,k} &=\frac{U}{4} \left(1+\cos(\theta)\right)-J z \cos^2(2\phi) \sin^2(\theta)\nonumber \\
&\quad\quad+\frac{\epsilon_k}{2}\left[1+\cos(\theta)\right]
\label{eq:alpha:phi}\\
\beta_{\phi,k} &=-\frac{\epsilon_k}{2} \left[1+\cos(\theta)\right] e^{2 i \phi},
\label{eq:beta:phi} 
\end{align}
where $\epsilon_k=2 J (\cos(k_x)+\cos(k_y))$. The Hamiltonians for the amplitude and phase modes are diagonalized by Bogoliubov transformations, as described in Appendix~\ref{app:Bogoliubov}, and the resulting spectra are presented in Fig.~\ref{fig:dispersion}. The general features of these spectra are as follows: (1) The amplitude mode is gapped everywhere in the phase diagram except at the QCP where it becomes gapless. (2) On the other hand, in the superfluid phase the phase mode corresponds to the Goldstone mode and is therefore gapless, while in the Mott phase the phase mode also becomes gapped. Furthermore, the (phase) amplitude mode becomes the (anti-)symmetric combination of the particle and hole excitations in the Mott phase, and since we are working in a particle-hole symmetric model their spectra become identical. (3) For all spectra shown in Fig.~\ref{fig:dispersion}, there is the large density of states near $U/2$ coming from high momentum modes, which will play an important role in fast dynamics. We note that the enumerated features of these spectra will be preserved for the case of the Bose-Hubbard model, although the spectra will change quantitatively.

\subsubsection{Wave function}
For each mode $\sigma\in\{\alpha,\phi\}$ and momentum $k$, the corresponding Hamiltonian is quadratic and can be solved by a Bogoliubov transform. Therefore, in terms of $b_{\sigma,k}$ operators the ground state for each $\sigma$ and $k$ is a squeezed coherent state. Since the various modes and momenta are non-interacting, we can write the total wave function (in terms of Holstein-Primakoff operators) as a product of squeezed states: 
\begin{align}
|\Psi\rangle&=\prod_{\sigma\in \{\alpha, \phi\},\,k>0} |\psi_{\sigma,k}\rangle\\
|\psi_{\sigma,k}\rangle&=
\sqrt{1-|c_{\sigma,k}|^2} e^{i \zeta_{\sigma,k}} e^{c_{\sigma,k} b^\dagger_{\sigma,k} b^\dagger_{\sigma,-k}}|0\rangle,
\label{eq:squeezedPsi}
\end{align}
where $c_{\sigma,k}$ is the squeezing parameter, $\zeta_{\sigma,k}$ is an additional phase, and $|0\rangle$ corresponds to the vacuum of Holstein-Primakoff (i.e. $|0\rangle=\prod_i b_{0,i}^\dagger |\text{empty}\rangle$). The phase $\zeta_{\sigma,k}$ contributes to the overall phase of the wave function and is therefore unimportant. The equilibrium value for the wave function parameter $c_{\sigma,k}$ is listed in Appendix~\ref{app:equilibriumMFpG}. In the same Appendix, we obtain expressions for defect density and the correlation functions in equilibrium, the properties of which are the subject of the next section.

\subsubsection{Schr\"odinger equation of motion}
During parametric tuning, the Hamiltonian given by Eq.~\eqref{eq:Heff2} remains quadratic. As a result, the wave function at all times may be written in the squeezed form of Eq.~\eqref{eq:squeezedPsi}. Therefore to understand the evolution of the wave function, we need to study the evolution of the parameters $c_{\sigma,k}(t)$ and $\zeta_{\sigma,k}(t)$. The evolution of the wave function in each mode is governed by the Schrodinger equation
\begin{align}
i\hbar &\left[\partial_t - 2A_{\sigma,k}^*
\left(b^\dagger_{\sigma,k}b_{\sigma,k}+b^\dagger_{\sigma,-k}b_{\sigma,-k}\right)\right]
|\psi_{\sigma,k}\rangle \nonumber\\
&\quad= H_{\sigma,k}(t)  |\psi_{\sigma,k}\rangle.
\label{eq:Schrodinger}
\end{align}
Here, the second term in the square brackets is a Berry phase term that is associated with the fact that the operator $b_{\sigma,k}$ itself transforms in time as the mean field evolves. The Berry phase $A(t)$ can be obtained by looking at the action of the coordinate transformation $M(t)$ from Eq.~\eqref{eq:M} on the vector $\vec{b}_i(t)=\left(b_{0,i}(t),b_{\alpha,i}(t),b_{\phi,i}(t)\right)$: 
\begin{align}
\vec{b}_i(t) \rightarrow & \vec{b}_i(t+\delta)= M(t+\delta) \cdot \vec{t}_i(t) \\
&= M(t+\delta) \cdot M^{-1}(t) \cdot \vec{b}_i(t)\\
&=\left[M(t)+\delta \partial_t M(t)\right] \cdot M^{-1}(t) \cdot \vec{b}_i(t),
\end{align}
where $\vec{t}(t)=\left(t_{0,i}, t_{+,i}, t_{-,i}\right)$. Thus we obtain the Berry phase
\begin{align}
A&=\frac{\vec{b}(t+\delta)-\vec{b}(t)}{\delta}=\left[\partial_t M(t) \right] \cdot M^{-1}(t) \\
&=\left(\begin{array}{c c c}
i \left(\dot{\chi}+\cos(\theta) \dot{\phi}\right) & -\frac{1}{2} \dot{\theta} + i \sin(\theta) \dot{\phi} & 0 \\
\frac{1}{2} \dot{\theta} + i \sin(\theta) \dot{\phi} &  i\left(\dot{\chi}- \cos(\theta) \dot{\phi}\right) & 0 \\
0 & 0 & i\left(\dot{\chi}-\dot{\phi}\right)
\end{array}\right).
\label{eq:BPh}
\end{align}
Setting $\dot{\chi}=-\cos(\theta)\dot{\phi}$, we remove the Berry phase for the $b_0$ operator and thus allowing us to set $b_0^\dagger b_0^\dagger \rightarrow 1$ in Eqs.~\eqref{eq:Salpha} and \eqref{eq:Sphi}. Next, by the assumption that the mean field is uniform, we drop the off-diagonal terms in $A$ that involve modes with finite $k$. Thus for finite $k$, the non-zero terms of the Berry phase are: $A_{\alpha,\alpha}= - 2 i \cos(\theta) \dot{\phi}$ and $A_{\phi, \phi}=-i (\cos(\theta)+1)\dot{\phi}$. Using this Berry phase, we find the evolution equation for the squeezing parameter
\begin{align}
i\hbar (\partial_t-2 A_{\sigma,\sigma}) c_{\sigma,k}=2 \alpha_{\sigma,k} c_{\sigma,k} + \beta_{\sigma,k} +\beta_{\sigma,k}^* c^2_{\sigma,k}
\label{eq:EOMBF}
\end{align}

In the above paragraph, we have focused on the dynamics of finite $k$ modes. Now we re-examine the dynamics of the $k=0$ mode, described by the vector $b_{0,i}$ of Eq.~\eqref{eq:M}. The time evolution of the direction of $b_{0,i}$ is described by the mean field equations of motion Eqs.~\eqref{eq:MFT1} and \eqref{eq:MFT2}. As we now demonstrate, the mean field equations of motion can be expressed in terms of $b_{0,i}$, $b_{\alpha,i}$, and $b_{\phi,i}$ operators. 

Consider the equation of motion Eq.~\ref{eq:Schrodinger} for the $k=0$ component of $b_{0,i}$.
The right hand side of Eq.~\ref{eq:Schrodinger} involves the Hamiltonian. For the k=0 case, the most significant contribution to the Hamiltonian comes from the terms that are first order in $b_{\alpha,i}$ and $b_{\phi,i}$ operators (which vanished for $k\neq0$ case by assumption of a uniform mean field):
\begin{align}
&H_\text{uniform}=-\frac{U}{2}\sum_{i} \sin(\theta) \left(b^\dagger_{\alpha,i} b_{0,i}+b^\dagger_{0,i} b_{\alpha,i}\right) \nonumber\\
&\quad\quad + J z \sum_{i} \left[\cos^2(2\phi) \sin(2\theta) \left(b^\dagger_{\alpha,i} b_{0,i}+b^\dagger_{0,i} b_{\alpha,i}\right)\right. \nonumber\\
&\quad\quad\quad\quad \left. +i\sin(4\phi) \sin(\theta) \left(b^\dagger_{\alpha,i} b_{0,i}-b^\dagger_{0,i} b_{\alpha,i}\right)\right],
\end{align}
where we used $b^\dagger_{0,i} b_{0,i} \rightarrow 1$. On the left hand side of Eq.~\ref{eq:Schrodinger}, we find that $b_{0,i}$ is coupled with $b_{\alpha,i}$ by a Berry phase Eq.~\eqref{eq:BPh}. This Berry phase is associated with the time evolution of the direction of $b_{0,i}$, and is precisely canceled by the Hamiltonian on the right hand side if the mean field equations of motion Eqs.~\eqref{eq:MFT1} and \eqref{eq:MFT2} are satisfied.

An alternative view of time evolution can be gained by noting the fact that the Hamiltonian is always quadratic means that the Heisenberg equations of motions for the quadratures $\langle b^\dagger_{\sigma, k} b_{\sigma, k} + b^\dagger_{\sigma, -k} b_{\sigma, -k}\rangle$, $\langle b^\dagger_{\sigma, k} b^\dagger_{\sigma, -k} \rangle$, and $\langle b_{\sigma, k} b_{\sigma, -k}\rangle$ close on themselves (see also Ref.~\cite{Lamacraft2007, Lamacraft2012} for $S=1$ superfluid tuning). That is, the equations of motion do not involve higher order terms. This fact can be exploited to study the evolution of any state that can be described by these three quadratures, including not only the ground state but also the thermal state. In fact, the evolution of the ground state using the quadratures method exactly matches the Schr\"odinger equation method Eq.~\eqref{eq:Schrodinger}. We summarize the quadrature method in Appendix~\ref{app:quads}.

\subsubsection{Defect density operator}
As we have already stated, one of the most experimentally useful observables is the defect density. In the truncated quantum rotor model, the defect density operator corresponds to $\left(S^z_i\right)^2$, and its expectation value, up to quadratic order may be written in the form
\begin{align}
\langle P_d \rangle =&\frac{1}{2}\left[1-\cos(\theta)\right] + \cos(\theta) \sum_k \langle b^\dagger_{\alpha k} b_{\alpha k} \rangle \nonumber\\
&\quad\quad\quad\quad\quad+\frac{1}{2} \left[1+\cos(\theta)\right] \sum_k \langle b^\dagger_{\phi,k} b_{\phi,k} \rangle.
\end{align}
In terms of the squeezing parameter of Eq.~\eqref{eq:squeezedPsi}, the defect density is 
\begin{align}
\langle P_d \rangle =&\frac{1}{2}\left[1-\cos(\theta)\right] + \cos(\theta) \sum_k \frac{|c_{\alpha k}|^2}{1-|c_{\alpha k}|^2} \nonumber \\
&\quad\quad\quad\quad+\frac{1}{2} \left[1+\cos(\theta)\right] \sum_k \frac{|c_{\phi k}|^2}{1-|c_{\phi k}|^2}.
\end{align}
In Appendix~\ref{app:equilibriumMFpG}, we provide details on how to compute the defect density, as well as the various correlation functions from Table~\ref{table:Measure} in equilibrium. The results of the equilibrium calculations are presented in section~\ref{sec:equilibrium}. The expression for the defect density is again used in conjunction with the Schr\"odinger equation of motion of the previous subsection to find the evolution of the number of defects following a ramp of the Hamiltonian parameters. The results of dynamics calculations are presented in section~\ref{sec:dynamics}.

\section{Results in equilibrium}
\label{sec:equilibrium}
All of the proposed methods (MF, MF+G, CMF, and ED) have advantages and disadvantages. The goal of this section is to establish these advantages and disadvantages in the simple context of a homogeneous system in equilibrium. As a confidence building measure, we try out the various approximate methods on both the Bose-Hubbard model as well as the quantum rotor model. In particular, we will compare the location of the phase boundary obtained using the various approximate methods to the one predicted by quantum Monte Carlo~\cite{Capogrosso-Sansone2007}. 

In addition to building our confidence, we will find some simple, yet useful, expressions for experimentally measured quantities. In particular, we obtain some expressions for defect density as well as the correlation functions listed in Table~\ref{table:Measure}. Defect statistics are of particular interest because they are accessible in current experiments~\cite{Bakr2009, Bakr2010, Sherson2010}. In addition, in Mott insulators defects, i.e. sites containing too many or too few bosons, roughly correspond to quasiparticle and quasihole excitations of the systems. One may expect (falsely) that upon crossing the phase boundary from the SF to the MI, the density of defects must decrease sharply. However, defect density is largely a local property of the system, and like other local properties does not show any significant structure near the phase transition. Using CMF and ED calculations we will show that nothing dramatic happens to the density of defects, nor short-range correlation functions listed in Table~\ref{table:Measure} at the transition point. However, the non-local physics must indeed show a diverging length-scale near the phase transition point, which we identify using MF+G method. Detection of this diverging length-scale could be a good way to locate the phase transition experimentally, therefore we derive simple closed form expressions for the correlation functions. 

This section is organized as follows. First, we compare the location of the SF-MI phase boundary obtained by various methods with its true location as obtained by quantum Monte Carlo simulations. Next, we compute the defect density using the various methods. Finally, we move on to compute the $g_2$ correlation functions. 

\subsection{Phase boundary}
A first test of the quality of the approximation is the location of the SF-MI phase boundary. Amongst methods that we are investigating, MF and CMF methods are able to find the phase boundary. The location of the phase boundary that can be found using MF+G method will coincide with the MF method, as we do not implement feedback of fluctuations onto the order parameter. As a standard, we compare the results of the various methods to those obtained from Monte Carlo. We note that all MF methods will generically tend to overestimate the importance of the ordered state, which in this case is the superfluid state. The reason for this tendency is that mean field methods do a poor job of taking into account the effects of long wave-length fluctuations that tend to destroy the emerging ordering. This problem is especially exacerbated in lower dimensions, where fluctuations are most important. 

We start by looking at the Bose-Hubbard model in two dimensions. In Fig.~\ref{fig:PhaseBoundary}a, we compare the QMC result~\cite{Capogrosso-Sansone2007} with MF and CMF results for various cluster sizes. We see that the CMF method approaches the QMC result as clusters get larger. Next, we plot the phase boundary for the Spin-1 model [see Fig.~\ref{fig:PhaseBoundary}b]. Again we see that the extent of the Mott phase increases as the cluster size increases. However, in both cases the extent of the Mott phase begins to saturate for 3x4 clusters, which are the largest clusters that we simulated.

In conclusion, the CMF method provides reasonable, but not quantitatively exact, estimates of the phase boundary. These estimates are significantly better for larger clusters. On the other hand, the qualitative features of  the phase boundary are quite reasonable. Therefore, we suggest that using CMF method (or ED method with equivalent size) to study dynamics cannot provide accurate answers, but it can be used for  qualitative answers. Further, as the location of the phase boundary depends on cluster size, one should adjust $J/U$ to compensate, especially if one wishes to study dynamics in the vicinity of the transition. That is, instead of using $J/U$ as the tuning parameter one should use $(J/U)/(J/U)_c$.

\subsection{Defect density}
\begin{figure}[h]
\includegraphics[height=4.3cm]{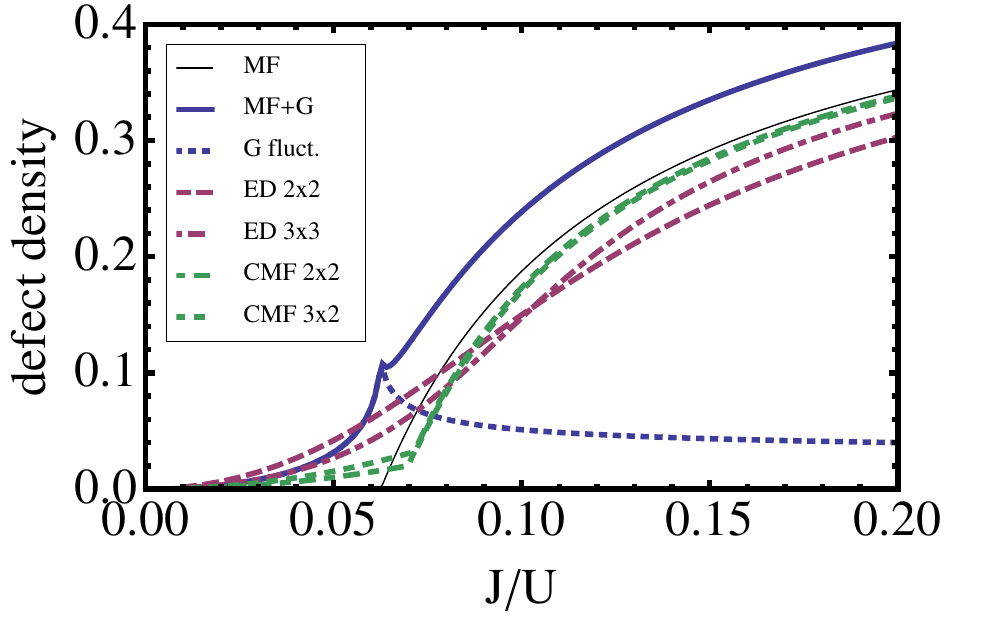}
\caption{Defect density in the Spin-1 model as a function of $J/U$ along the particle hole symmetric line ($\mu=0.5$). CMF and ED methods converge away from the phase transition as the system/cluster size is increased. On the other hand, close to the phase transition the convergence is much worse. The MF+G method seems to always overestimate the defect density (as compared to the converged result which should lie between the 3x2 CMF and the 3x3 ED). The poor performance of the MF+G method is likely due to the lack of self-consistency.}
\label{Fig:defectDensity}
\end{figure}

Defect density, as defined in Table~\ref{table:Measure} is especially interesting as it is measured experimentally.  Currently, experiments can only distinguish whether the number of bosons on a particular site is even or odd so instead of the full counting statistics what is available is the density  of ``defects." 

In Fig.~\ref{Fig:defectDensity}, we plot the defect density as a function of $J/U$ along the particle-hole symmetric line for the case of the Spin-1 model for ED, MF, CMF, and MF+G methods. Comparing CMF and ED results with different cluster/system sizes, we see that the the ED and CMF results seem to converge as the cluster/system size increases. The convergence is good everywhere except in the vicinity of the phase transition. The lack of convergence in the vicinity of the phase transition is related to the importance of long wavelength fluctuations which cannot be captured by small cluster/systems size methods.

An important question is the qualitative behavior of the defect density near the transition point. Here, the ED method shows  the defect density monotonically increasing with $J/U$ without any singular features in the vicinity of the phase transition (since ED works on finite sized systems, this lack of singular behavior is expected). On the other hand MF and CMF methods also show the defect density monotonically increasing with $J/U$ but with a kink at the phase transition point, while the MF+G shows a slight bump in the vicinity of the phase transition. In the CMF method we see that the kink becomes smoother as the size of the cluster is increased. We argue that the defect density is mostly a short wavelength phenomenon and therefore should be only weakly sensitive to long wavelength fluctuations. On the other hand, the kink/bump is associated with long wavelength fluctuations, which are given excess importance within mean field theory description. Thus, we expect that a kink, but probably not a bump, can be found in the vicinity of the phase transition. However, as suggested by CMF with larger cluster sizes, this feature could be rather subtle, and thus it is a poor way of detecting the transition experimentally. 

\subsection{Correlation functions}
\begin{figure*}
\begin{minipage}[b]{0.2cm}
	{\bf (a)}
	
	\vspace{2.8cm}
\end{minipage}
\begin{minipage}[t]{5.6cm}
	\includegraphics[scale=0.55]{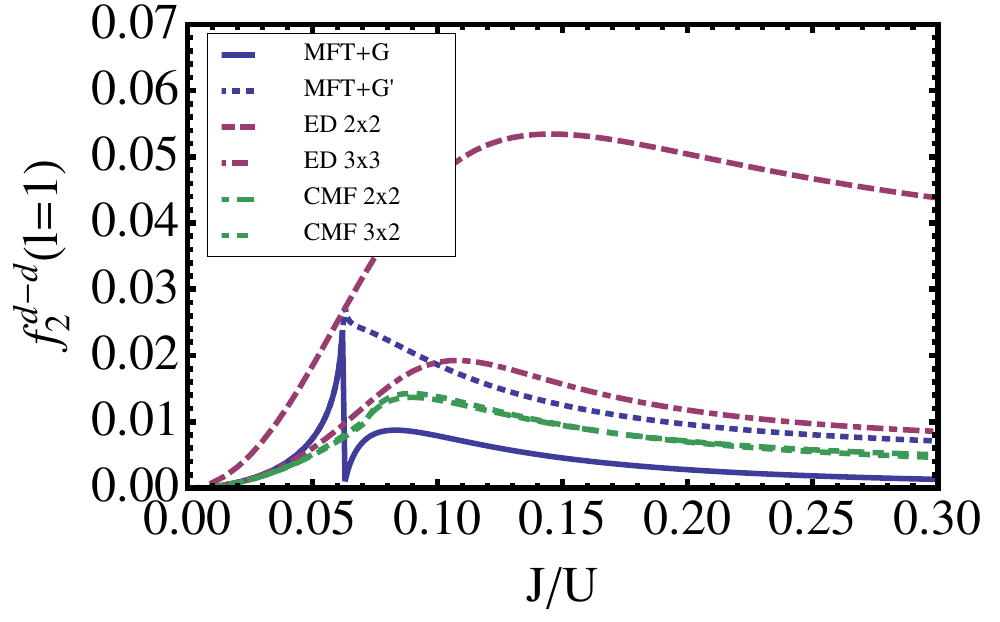}
\end{minipage}
\begin{minipage}[b]{0.2cm}
	{\bf (b)}
	
	\vspace{2.8cm}
\end{minipage}
\begin{minipage}[t]{5.6cm}
	\includegraphics[scale=0.55]{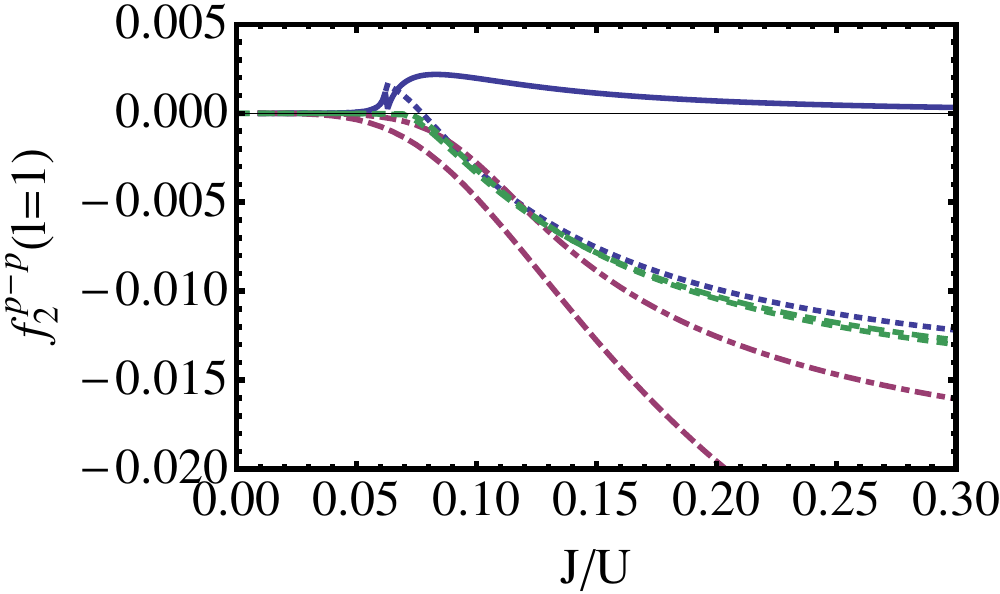}
\end{minipage}
\begin{minipage}[b]{0.2cm}
	{\bf (c)}
	
	\vspace{2.8cm}
\end{minipage}
\begin{minipage}[t]{5.6cm}
	\includegraphics[scale=0.55]{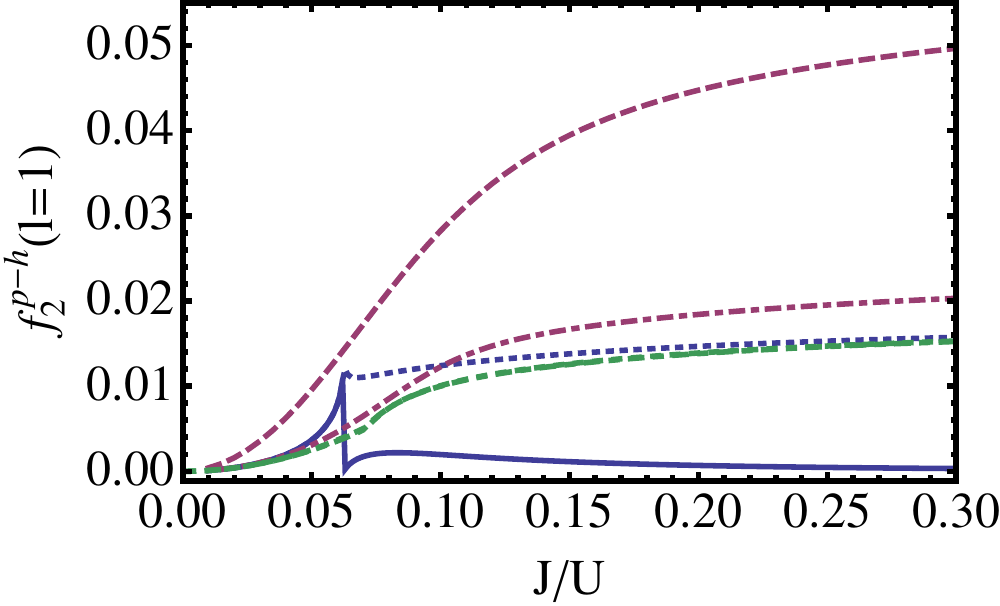}
\end{minipage}
\caption{(a) Correlation functions (a) $f_2^{d-d}(l=1)\equiv \langle P_d(i+1) P_d(i)\rangle-\langle P_d(i+1) \rangle\langle P_d(i) \rangle$, (b) $f_2^{p-p}(l=1)=\langle P_p(I+1) P_p(i)\rangle-\langle P_p(i+1) \rangle\langle P_p(i) \rangle$ and (c) $f_2^{p-h}(l=1)=\langle P_p(i+1) P_h(i)\rangle-\langle P_p(i+1) \rangle \langle P_h(i) \rangle$ as a function of $J/U$, computed using Mean Field with Gaussian fluctuations, Exact Diagonalization, and Cluster Mean Field. For the curve labeled MF+G, we used quadratic order on the Superfluid side and quartic on the Mott side; for MF+G' we used quartic order on both sides. The MF+G curve shows a jump indicating the importance of quartic fluctuations for computing nearest neighbor correlations. Overall, we see that all methods show qualitative similarities, and tend to converge as cluster size/system size is increased. The biggest differences occur in the vicinity of the phase transition where MF+G, MF+G' and CMF methods show a kink and exact diagonalization does not. }
\label{Fig:f2l1}
\end{figure*}

In this subsection we investigate the short range (nearest neighbor) correlation functions using CMF, ED, and MF+G methods as well as long range correlation functions using the MF+G method. Our goal is to benchmark these methods as well as to establish some relatively simple results which could be of use for experimental data fitting.

We begin by looking at the defect-defect correlation function
\begin{align}
f_2^{d-d}(l)\equiv \langle P_d(i+l) P_d(i)\rangle-\langle P_d(i+l) \rangle \langle P_d(i) \rangle.
\end{align}
Using ED, CMF, and MF+G methods we shall compute the nearest-neighbor defect-defect correlation function $f_2^{d-d}(l=1)\equiv \langle P_d(i+1) P_d(i)\rangle-\langle P_d(i+1) \rangle \langle P_d(i) \rangle$ and compare the results. We expect that deep in the Mott insulator, there are no defects, and therefore $f_2^{d-d}(l=1)\rightarrow0$. On the other hand, as we move into the superfluid the bosons become weakly interacting and therefore the defect-defect correlations decrease (there is a subtle point that for the case of the Spin-1 model particle-particle and particle-hole correlations persist, even in the non-interacting case $J/U\rightarrow \infty$). In the vicinity of the transition, the interactions are strong and the defect density is large, thus we expect a maximum in $f_2^{d-d}(l=1)$. Indeed, plotting  $f_2^{d-d}(l=1)$ [see Fig.~\ref{Fig:f2l1}a], we find that all methods agree qualitatively. Quantitatively, we also find reasonable agreement with two exceptions. First, the $2\times2$ ED method does a particularly poor job, as nearest neighbors are doubly connected to each other -- once by the direct bond and a second time by the periodic boundary condition. Second, the MF+G method shows a jump at the phase transition, the CMF method shows a kink, and the ED shows a smooth curve. The origins of the disagreement at the phase transition are similar to those stated for the defect density, with one exception. The jump in the MF+G method is somewhat artificial, as we have used quadratic order expression on the superfluid side but are forced to resort to quartic order expression on the Mott side as quadratic order expression becomes zero, see Appendix~\ref{app:CF}. Indeed, the jump is replaced by a kink if we use quartic order expression on both sides. 

To gain better understanding of the defect-defect correlation function, we look at its constituents: 
\begin{align}
f_2^{d-d} = f_2^{p-p} + f_2^{p-h} + 2 f_2^{p-h}
\end{align}
the particle-particle (or equivalently, at the particle-hole symmetric point, the hole-hole) correlation function $f_2^{p-p}(l=1)\equiv \langle P_p(i+1) P_p(i)\rangle-\langle P_p(i+1) \rangle \langle P_p(i) \rangle$ [Fig.~\ref{Fig:f2l1}b] and the particle-hole correlation function $f_2^{p-h}(l=1)\equiv \langle P_p(i+1) P_h(i)\rangle-\langle P_p(i+1) \rangle\langle P_h(i) \rangle$ [Fig.~\ref{Fig:f2l1}c]. The particle-particle correlation function captures the fact that particles tend to repel each other, and thus we expect it to be negative (this remains true even in the non-interacting case due to the hopping term preferring particles next to holes). On the other hand, we expect that the particle-hole correlation function remains positive throughout the phase diagram, capturing the physics of the tunneling term which tends to form particle-hole fluctuations on nearest neighbor sites. Using ED and CMF methods, we find the expected trends in both $f_2^{p-p}(l=1)$ and $f_2^{p-h}(l=1)$. On the other hand, we find that the MF+G method predicts a region of $J/U$ where $f_2^{p-p}(l=1)>0$. We attribute this failure to the fact that the MF+G method only captures single quasiparticle physics, and thus is unable to capture quasiparticle-quasiparticle repulsion.

Quantitatively, for both $f_2^{p-p}(l=1)$ and $f_2^{p-h}(l=1)$ we see that the CMF and the ED method seem to converge as cluster/system size is increased. As in the case of defect-defect correlations, $2\times2$ ED does a particularly poor job due to the nearest neighbors being doubly connected. Near the transition, the particle-hole correlations are largely responsible defect-defect correlations, thus $f_2^{p-h}(l=1)$ [Fig.~\ref{Fig:f2l1}c] is very similar to $f_2^{d-d}(l=1)$ [Fig.~\ref{Fig:f2l1}a] (up to a factor of two due to the definitions) and all three methods ED, CMF and MF+G give reasonable results. Finally, we mention that we present the normalized versions of these correlation function, e.g. $g_2^{d-d}(l=1)=f_2^{d-d}(l=1)/\langle P_d\rangle^2+1$, in Appendix~\ref{app:CFN}.

\begin{figure}[h]
\includegraphics[width=8.5cm]{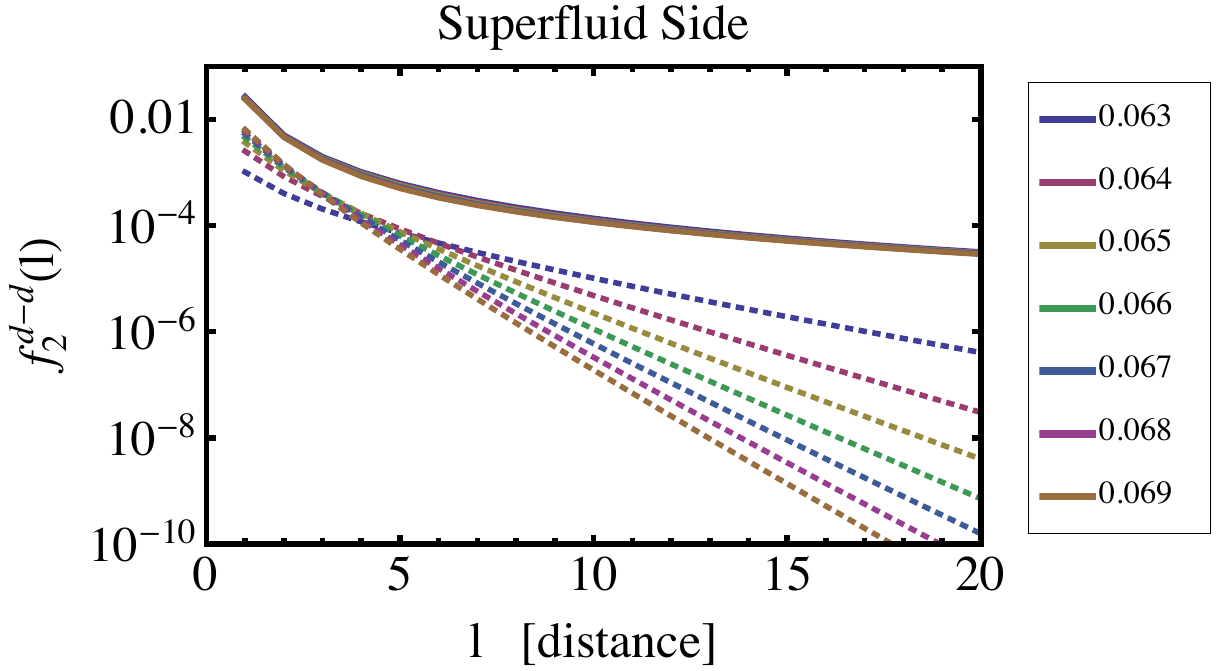}\hspace{1cm}
\\
\vspace{0.4cm}
\includegraphics[width=8.5cm]{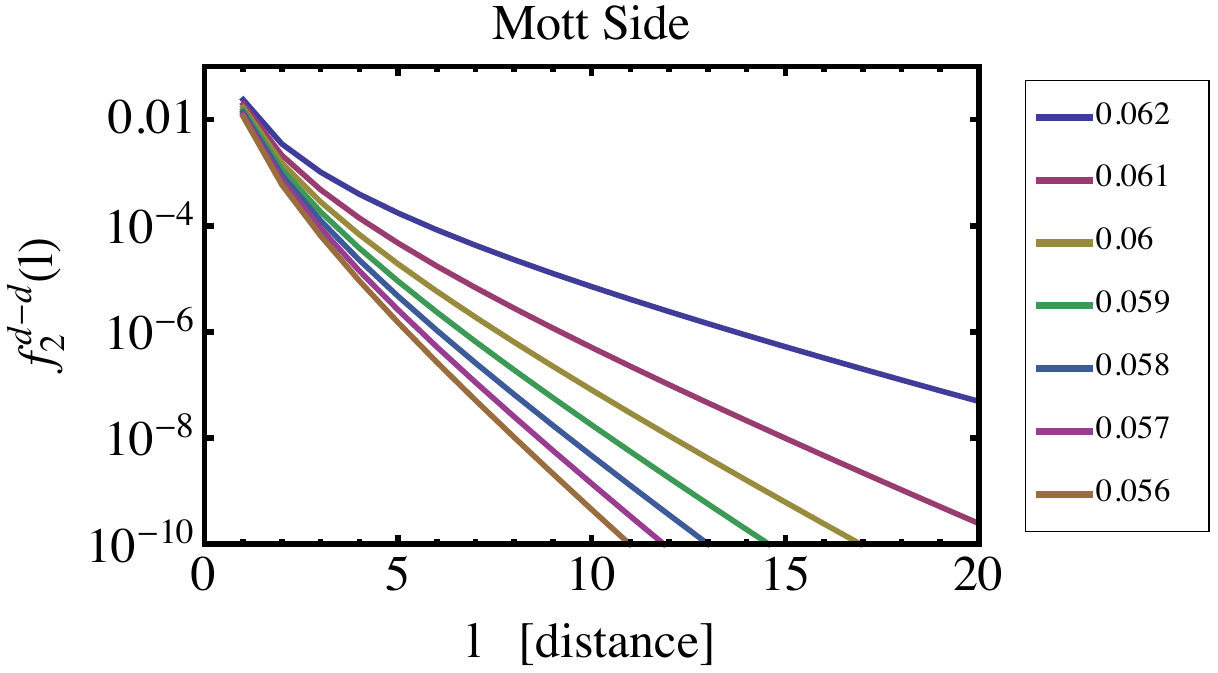}
\caption{Decay of the correlation function $f^{d-d}_2(l)=\langle P_d(i+l) P_d(i)\rangle-\langle P_d(i+l) \rangle\langle P_d(i) \rangle$ as a function of separation distance $l$ plotted on a semi-log scale for various values of $J/U$ in the vicinity of the transition $(J/U)_c=0.0625$. On the superfluid side, we plot all contributions to $g_2(l)-1$ including both second and fourth order terms with solid lines, and the part from second order term only with dotted lines. On the Mott side, the second order term vanishes, and we only plot the contributions from the fourth order terms. See text for details.}
\label{Fig:decay}
\end{figure}

Having verified that the MF+G method provides results similar to ED and CMF for correlations on nearest neighbor sites, we move on to the question of how correlations decay at large distances. This question we attack using only the MF+G method, as the other two methods are not easily extended to the long distances. Specifically, we look at the decay of defect-defect correlations $f^{d-d}_2(l)$ in the vicinity of the transition, on both the superfluid and the Mott insulating side, Fig.~\ref{Fig:decay}. To compute the correlator we have used both the second order (in fluctuations) term given by Eq.~\ref{Eq:fdd2} and the fourth order term given by Eq.~\ref{Eq:f24}. 

On the superfluid side, both the second and the fourth order terms are non-zero. Naively, since the second order term is non-zero, we may consider stopping at this order. However, because the second order term comes from fluctuations of amplitude modes only, while the fourth order term has contributions from both amplitude and phase modes, it is important to include the fourth order term. The reason why phase modes become important is that their gap closes at small momenta, while amplitude modes are always gapped. As a result, the contributions from amplitude modes show exponential decay with distance, as can be seen by plotting the second order term: $f^{d-d (2)}_2(l)\sim \exp(-l/\xi)$ (dotted traces in Fig.~\ref{Fig:decay}). Moreover, as we approach the transition, the length scale $\xi$ diverges (dotted traces become flatter in Fig.~\ref{Fig:decay}). On the other hand, contributions from phase modes show power law decay with distance, which becomes dominant when plotting the sum of second and fourth order terms: $f^{d-d}_2(l)\sim f^{d-d (2)}_2(l)+f^{d-d (4)}_2(l)\sim l^{-2}$ (solid traces in Fig.~\ref{Fig:decay}). We note that if we were to proceeding to higher orders, we would expect to find more negative power-laws, and thus we stop at the fourth order.

The Mott insulating side is more straightforward for two reasons: (1) the second order term vanishes, and (2) both phase and amplitude modes become everywhere gapped, thus $f^{d-d}_2(l)\sim \exp(-l/\xi)$, Fig.~\ref{Fig:decay}. Again, the length scale $xi$ can be seen diverging as one approaches the transitions as the traces in Fig.~\ref{Fig:decay} become flatter.

\begin{figure}[h]
\includegraphics[width=6.5cm]{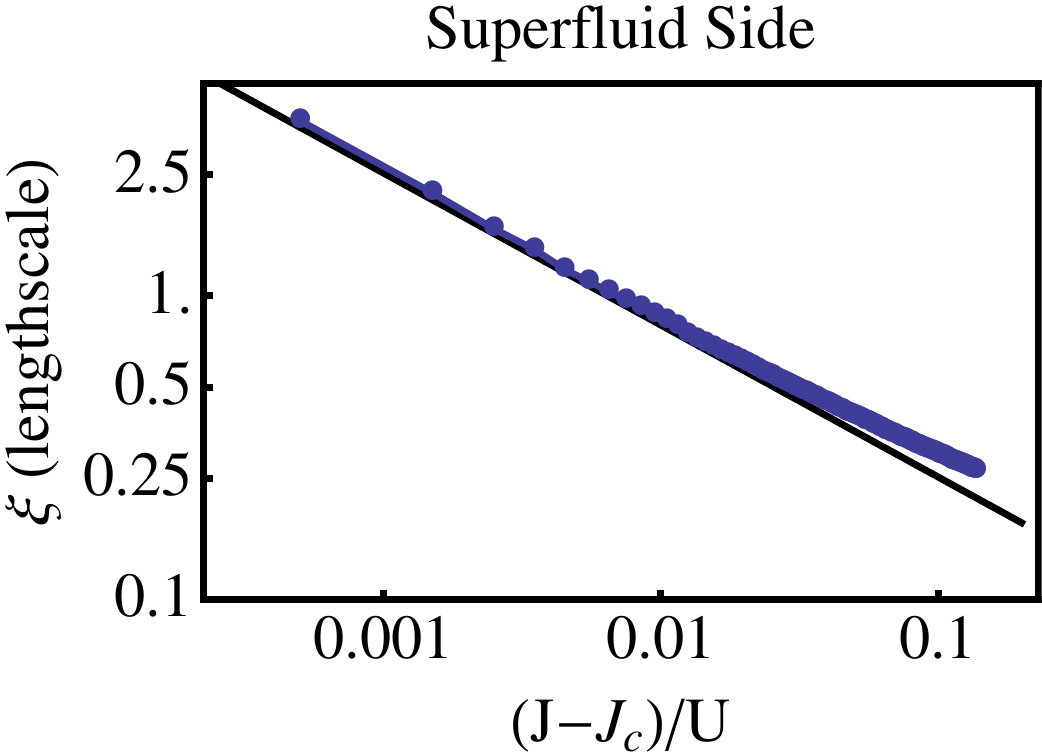}\hspace{1cm}
\\
\vspace{0.4cm}
\includegraphics[width=6.5cm]{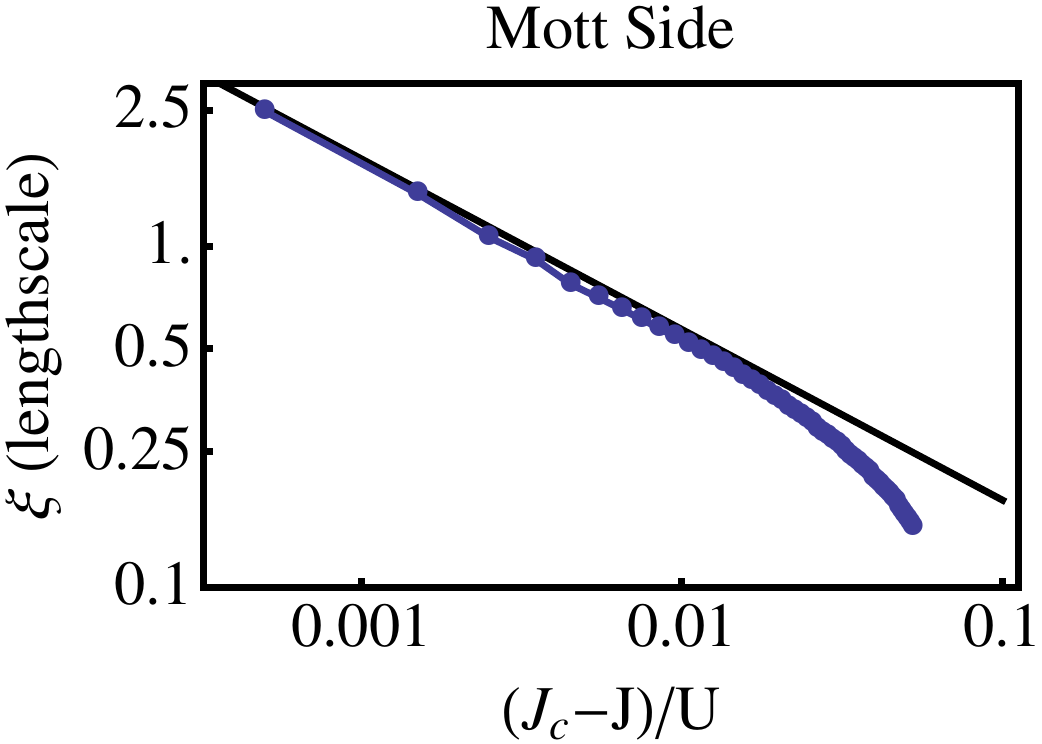}
\caption{Diverging coherence length as a function of distance to the phase transition on a log-log plot. The scaling is consistent with MFT expectations of $\nu=0.5$. }
\label{Fig:scaling}
\end{figure}

We can extract the correlation length scale $\xi$, that diverges at the phase transition, by fitting $f^{d-d}(2)_2(l)$ on the superfluid side and $f^{d-d}(4)_2(l)$ on the Mott side with the form $\sim\exp(-l/\xi)$,  Fig.~\ref{Fig:scaling}. We find that the scaling of $\xi \sim |r|^{-\nu}$ is consistent with MFT result $\nu=0.5$, where $r=J/U-(J/U)_c$ is the detuning away from the QCP. Using these notions, we obtain an approximate expression for $f^{d-d}_2$ by fitting  Fig.~\ref{Fig:decay}, which is valid near the phase transition, but is also reasonably accurate through the phase diagram
\begin{align}
f^{d-d}_{2,SF}(\ell) &\approx \frac{0.013-0.18 \,r}{l^2}+\frac{2.54\, r}{l}e^{-11.29\cdot l \cdot r^{0.5}},\\
f^{d-d}_{2,MI}(\ell) &\approx \frac{0.031-1.54\, |r|}{l^2} e^{-16.31 \cdot l \cdot |r|^{0.5}}.
\label{eq:correlation}
\end{align}

\begin{figure*}
\includegraphics[width=\textwidth]{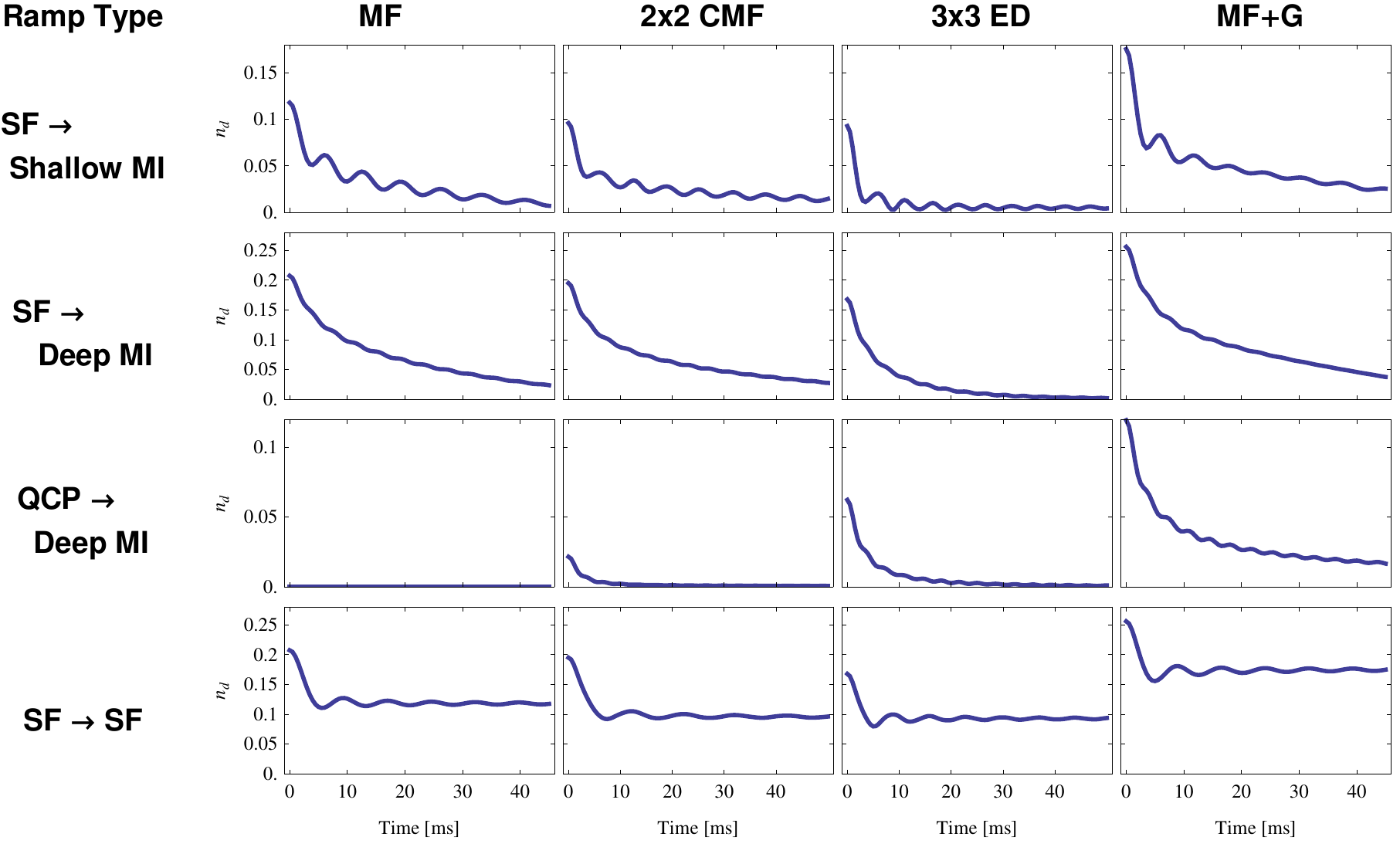}
\caption{Defect density (immediately following the end of the ramp) as a function of the ramp time. The calculations were performed for four different ramp types (as indicated by row headings), using the four different methods (as indicated by column headings). The parameters used in calculations correspond to Rb atoms in an optical lattice, see Appendix~\ref{app:exp} for details. Note that the value of the mean field (for the MF, CMF, and MF+G methods) for the QCP $\rightarrow$ deep MI ramp is identically zero throughout the ramp. As a result, for the MF method there are no defects, while for the CMF and MF+G methods  defects appear due to fluctuations. }
\label{fig:d1} 
\end{figure*}

\begin{table*}
\begin{tabular}{c||c}
Ramp Type & Functional Form\\
\hline
SF $\rightarrow$ Shallow MI & $V(t)=11+(16-11)t/t_\text{max}$\\
\hline
SF $\rightarrow$ Deep MI & $V(t)=10+(25-10)t/t_\text{max}$\\
\hline
QCP $\rightarrow$ Deep MI (MF, MF+G)& $V(t)=12.0315+(25-12.0315)t/t_\text{max}$ \\
QCP $\rightarrow$ Deep MI (CMF, ED)&\newline $V(t)=11.5949+(25-11.5949)t/t_\text{max}$\\
\hline
SF $\rightarrow$ SF & $V(t)=10+(11-10)t/t_\text{max}$
\end{tabular}
\caption{Functional forms of the optic lattice intensity as a function of time used for the various types of ramps. Note that in the ramps originating at the QCP, the functional form depends on the location of the QCP. For the case of ED, which lacks a QCP, the location of the QCP from CMF was used.}
\label{table:ramps}
\end{table*}

\begin{figure}
\includegraphics[width=\columnwidth]{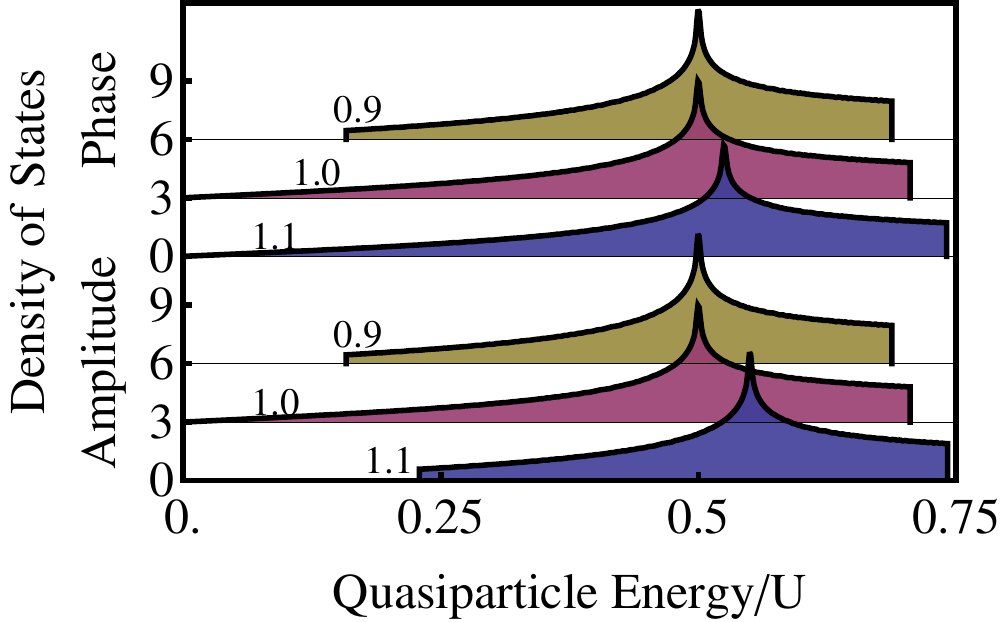}
\caption{Density of states of the phase (upper panel) and amplitude (lower panel) modes, as a function of the energy (measured in units of the interaction parameter $U$) for the Spin-1 model (from the MF+G method). The density of state curves are plotted for three values of $(J/U)/(J/U)_c$ as indicated by the number next to the curve (consecutive curves were displaced vertically for clarity). The values of $(J/U)/(J/U)_c$ correspond to the superfluid phase ($(J/U)/(J/U)_c=1.1$), the QCP ($(J/U)/(J/U)_c=1.0$), and the Mott phase ($(J/U)/(J/U)_c=0.9$). The density of state curves have several interesting features.
First, is the appearance of a gap in the amplitude mode on both sides of the QCP. Second, is the appearance of a gap in the phase mode on the Mott side but not the superfluid side of the QCP. Finally, the presence, throughout the phase diagram, of a logarithmic singularity in the density of states, associated with momenta near $k_x \pm k_y = \pm \pi$, in the vicinity of $\sim U/2$ for both amplitude and phase modes. We note that this singularity is associated with the fast periodic oscillations seen in dynamics [Fig.~\ref{fig:d1}].}
\label{fig:DOS}
\end{figure}

\begin{figure}
\includegraphics[width=\columnwidth]{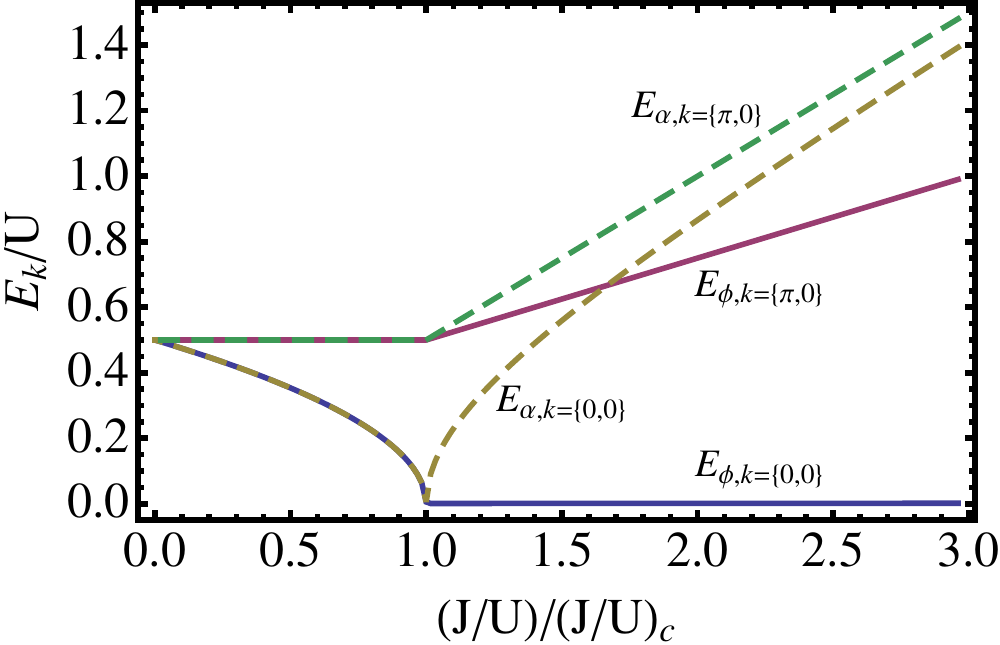}
\caption{Quasiparticle energy at the bottom of the phase and amplitude bands ($k=\{0,0\}$) and at the peak of the density of states ($k=\{0,\pi\}$, see Fig.~\ref{fig:DOS}) as a function of $J/U$ across the Mott insulator (LHS) to superfluid (RHS) phase transition for the Spin-1 model (from the MF+G method).}
\label{fig:band}
\end{figure}

\section{Dynamics of parametrically ramped systems}
\label{sec:dynamics}
The goal of this section is to understand the dynamics of the homogeneous spin-1 quantum rotor model undergoing a parametric ramp. That is, we consider the dynamics as we ramp the parameters $J$ and $U$ in time in the vicinity of the quantum critical point (QCP) separating the superfluid and the Mott insulator. We note that the $\mu$ term in Eq.~\eqref{eq:HSpin1} commutes with the other two terms, and therefore the tuning of the chemical potential has no effect on dynamics. 
We investigate dynamics across all timescales, from very fast ramps (timescale $1/U$) to very slow ramps (timescale $1/J$), with the goal of seeing the non-universal dynamics for fast ramps cross-over to critical scaling dynamics for slow ramps. We focus on two types of ramps (1) ramps that start in the superfluid phase and stop in the Mott insulating phase; and (2) ramps that start on the QCP and stop in the Mott insulating phase. The primary purpose of investigating the crossover into the universal scaling is to gain quantitative understanding of whether this regime can be observed experimentally. That is we want to obtain (1) a reasonable estimate for how slow one needs to ramp in order to be in the universal regime; and (2) an understanding of what observable to measure experimentally and whether it is detectable. Indeed, we will show that the universal scaling regime sets in for ramps of timescale $\sim 10/J$, and the density of defects (i.e.density of doubly occupied and empty sites) can be used as an experimental observable if the ramp goes sufficiently deeply into the Mott Insulator.

We remark that ramps we consider always go from the ordered (i.e. superfluid) to the disordered (i.e. Mott insulating phase) state, and thus can be thought of as opposite of the usual Kibble-Zurek process that describes the appearance of excitations and long range order~\cite{Kibble1976, Zurek1985} as the system is ramped from the disordered phase into the ordered phase (i.e. Mott insulator $\rightarrow$ SF). Ramps toward the disordered phase are technically easier to describe because they do not require modeling spontaneous symmetry breaking via Spinoidal decomposition, which falls outside the realm of mean field theories (i.e. MF, CMF, and MF+G). 

Explicitly, in a ramp towards the ordered phase collective modes of the system that are associated with ordering become unstable. In the language of susceptibilities, the susceptibility $\chi_q(t)$ associated with ordering
\begin{align}
\Delta_q(t)=\int dt' \chi_q(t-t') \Delta_q(t')
\end{align}
acquires a pole with positive imaginary frequency. Although this type of poles, with a positive imaginary frequency, appear in mean field theories there are no fluctuations to seed their growth. On the other hand, in the ramp towards the disordered state, both the order parameter dynamics and the dynamics of the other modes of the system are well defined within mean field theories (i.e. MF, CMF, and MF+G). The order parameter starts out finite and its dynamics correspond to the decay of its amplitude. Further, all of the normal modes of the system remain stable, and thus well described by the quadratic theory.

Before proceeding with the numerical calculations, we argue that the dynamics in the vicinity of the superfluid-Mott insulator QCP has two regimes: fast (non-universal) and slow (universal) regime. The fast regime, which has been the focus of the recent experiments~\cite{Bakr2010}, is associated with timescales comparable to the inverse of the bandwidth of the phase and amplitude modes, that is $t_\text{ramp}\sim2/U \sim1/ 2 J z $. In this regime dynamics is largely local and our CMF and ED approximations work well. On the other hand, the slow regime is associated with universal long-wavelength physics in the vicinity of the QCP, and timescales comparable to the inverse of the energy scale over which the dispersion is linear, i.e. $t_\text{ramp}\sim1/J$. Ramping through the Superfluid $\rightarrow$ Mott transition results in the creation of a number of excitations (phase and amplitude modes) in the system having a density $n_\text{ex}$. $n_\text{ex}$ is controlled by the ramp rate, and is believed to exhibit a universal power law in the slow regime
\begin{align}
n_\text{ex} \sim r^\frac{d \nu}{z \nu +1},
\end{align}
where $r=1/t_\text{ramp}$ is the ramp rate, $d$ is the dimensionality of the system, $\nu$ is the coherence length critical exponent, and $z$ is the dynamical critical exponent~\cite{Polkovnikov2011}. Plugging in the values of the critical exponents for the Superfluid-Mott transition at integer filling of the Bose-Hubbard model (i.e. at the particle-hole symmetric point) $d=2$, $z=1$, and $\nu=1/2$ we obtain $n_\text{ex} \sim r^{2/3}$ (at least using the mean field exponents -- which should correspond to our mean-field description of the QCP). These arguments will be made more precise as we analyze the numerical data.

To study the creation of excitations and defects upon ramping through a phase transition, we must specify the protocol for parametrically tuning $J$ and $U$. One important consideration is that only the density of defects (i.e. $\langle (S^z)^2 \rangle$), but not the density of single quasi-particle excitation (i.e. phase and amplitude modes of Eq.~\eqref{eq:Bogoliubov}), is experimentally accessible. Therefore, we will look at both the density of defects and the density of quasi-particle excitations. We note that the density of defects is not a constant of motion, and will fluctuate following the end of the ramp. Thus, for concreteness,  we choose to focus on the density of defects immediately following the end of the ramp. One way to make a direct connection between defects and excitations is to ramp the system deep into the Mott regime, where defects indeed correspond to excitations. Following this consideration, we shall consider both (1) shallow ramps that end in the Mott phase but close to the superfluid-Mott transition, and (2) deep ramps that end deep in the Mott phase (close to $J=0$).  Another important consideration, is the initial point for the start of the ramp. As we shall show, starting right at the phase transition seems to be advantageous for observing universal scaling behavior, as there are fewer fluctuations and the scaling sets in for shortest ramp times ($\sim 1/J$). However, experimentally preparing the system at the phase transition is a difficult task, due to the long equilibration times. A summary of ramp profiles that we investigate is provided in Table~\ref{table:ramps}, and the connection between the optical lattice intensity and the Hubbard parameters is provided in Appendix~\ref{app:exp}.

We begin by looking at the number of defects that are created by fast ramps with ramp times from zero to $50\,\text{ms} \sim 2\pi \hbar/J_c$. As we sweep through a range of values of $J$ and $U$, for definitiveness we shall alway compare time scales to the value of $J$ and $U$ at the critical point: $2 \pi \hbar/J_c=62\,\text{ms}$  and $2 \pi \hbar/U_c=3.88\,\text{ms}$. Naively, we would expect that in order to create or remove a defect we must move an atom from one site to another, which is associated with the tunneling time $2\pi \hbar/J_c$. In Fig.~\ref{fig:d1} we plot the number of defects created as a function of the ramp time, for various ramp profiles, calculated using the four different methods (MF, CMF, ED, and MF+G). Surprisingly, we see quite a lot of structure in the number of defects created even for ramp times as short as $4\,\text{ms} \sim 2 \pi \hbar/U_c$. The emergence of defects for such short timescales can be understood by considering a two site quantum rotor model with total $S^z=0$ (i.e. populated by two bosons). The spectrum of this model, for small $J$, consists of a lower branch (composed of mostly $|0,0\rangle$ state, with a small admixture of $|1,-1\rangle$ and $|-1,1\rangle$ states) and a pair of upper branches (composed mostly of $|1,-1\rangle$ and $|-1,1\rangle$ states with a small admixture of $|0,0\rangle$ state). The gap between the upper and lower branches is set by the on-site repulsion $U$. Hence, the important timescale for characterizing adiabaticity of the process is set by $2\pi \hbar/U_c$, and the dynamics will be non-adiabatic for ramps that are fast compared to this timescale. In the many-site system, there is no such gap, but there is a large density of states in the vicinity of $U/2$ scale [see Fig.~\ref{fig:DOS}]. The energy of this maximum, as a function of $(J/U)/(J/U)_c$ for both the amplitude and phase mode is plotted in Fig.~\ref{fig:band}, confirming that in the vicinity of the quantum critical point the maximum in the density of states is always near $U/2$. This maximum results in the appearance of oscillatory features at the corresponding timescales. 

One may wonder whether other features in the density of states are reflected in the dynamics. A particularly interesting feature is the gap that appears in the amplitude mode on either side of the phase transition. In particular, on the superfluid side, the amplitude mode is associated with the Higgs phenomenon, and the gap is related to the Higgs mass. We remark that we find no direct signatures of the Higgs mass (using all of our methods) in ramps that go all the way across the phase transition. However, the Higgs may be excited by performing small ramps or quenches on the superfluid side. Alternatively, the Higgs strongly couples to lattice modulations, and can be studied via lattice modulation spectroscopy as suggested in Ref.~\cite{Huber2007, Podolsky2011}. Experimental and theoretical investigations of the Higgs will be the subject of an upcoming manuscript~\cite{usHiggs}.

Thus far, our description of fast dynamics has not appealed to the presence of a phase transition. Indeed, because we are probing features at very short timescales and thus high energies, the properties of the phase transition, and in fact its very presence, are hard to see. As an example, we consider a set of ramps that stay completely on the superfluid side, see last row of Fig.~\ref{fig:d1}. We see that the timescales for the number of defects created qualitatively show features similar to ramps that cross the phase transition.

\begin{figure*}
(a)\!\!\!\!\!\!\includegraphics[width=5.5cm]{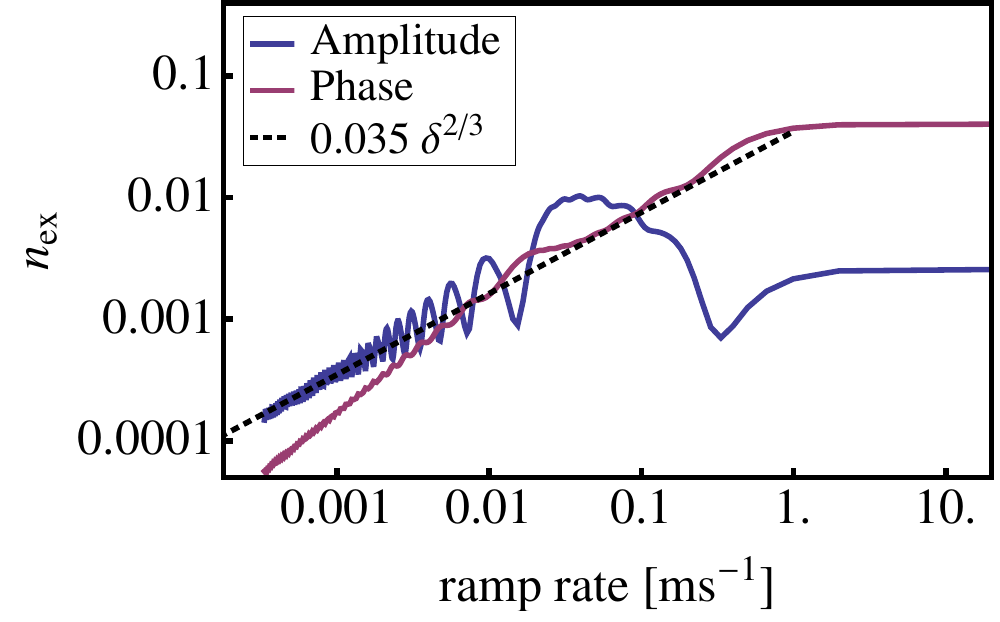}~~
(b)\!\!\!\!\!\!\includegraphics[width=5.5cm]{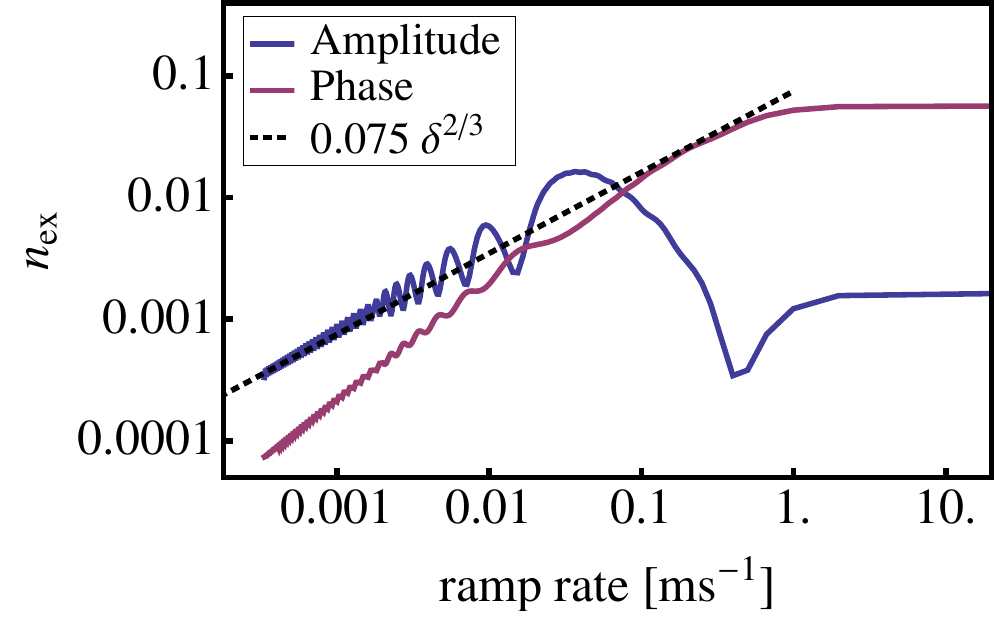}~~
(c)\!\!\!\!\!\!\includegraphics[width=5.5cm]{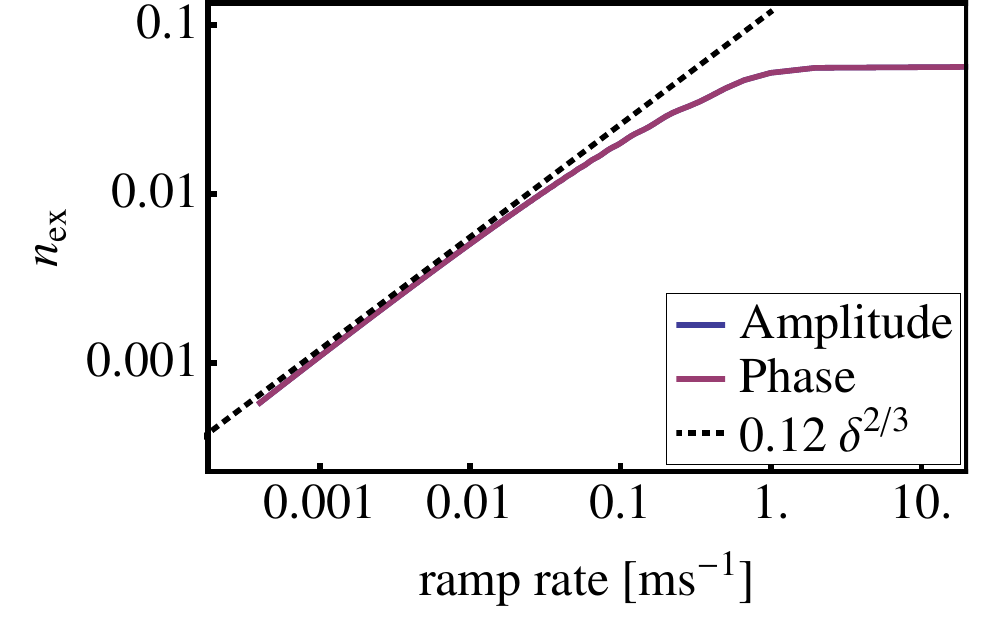}
\caption{Excitation density as a function of ramp rate ($1/t_\text{ramp}$) on a Log-Log plot. (a) Starting in SF ending in shallow MI. (b) Starting in SF ending in deep MI. (c) Starting at the critical point and ending deep in the MI (here, the MF order parameter is zero throughout). Note that starting at the QCP results in the observation of scaling at for shorter ramp times. }
\label{fig:d3}
\end{figure*}

Having understood fast timescales, we move on to the slow timescales. Here, we resort to using the MF+G method which is able to capture the long wavelength fluctuations. Since the MF+G method involves excitations of single particle modes, we can study both the number of excitation, as well as the number of defects created in the phase transition. In Fig.~\ref{fig:d3} we plot the number of single particle excitations created as a function of the ramp rate (inverse of ramp time), for various protocols.  We see that in all cases, for slow ramps, the amplitude mode displays the expected $2/3$ power law. Furthermore, with the exception of the protocol in which we start at the QCP, the majority of excitations are carried by the amplitude mode. However, timescales at which the scaling becomes easily discernible depend on the protocol.  For the two protocols starting in the superfluid, the excitation number exhibits some oscillations (with the period $\sim 2 \pi \hbar/J_c$ at criticality) before the power law can be observed. As a result, the power law only becomes apparent for relatively slow ramp rates, starting with $\sim 0.01 \text{ms}^{-1}\sim \frac{1}{2}(J_c / 2\pi \hbar)$. On the other hand for ramps that start at the QCP, the amplitude and phase modes are identical (amplitude [phase] mode corresponds to the symmetric [anti-symmetric] combination of particle and hole excitations of the Mott insulator), and the power law scaling sets in for relatively faster ramp rates $\sim 0.1 \text{ms}^{-1}\sim 6 (J_c / 2 \pi \hbar)$. Both of these timescales are comparable to what is possible in present experiments.

\begin{figure*}
(a)\!\!\!\!\!\!\includegraphics[width=5.5cm]{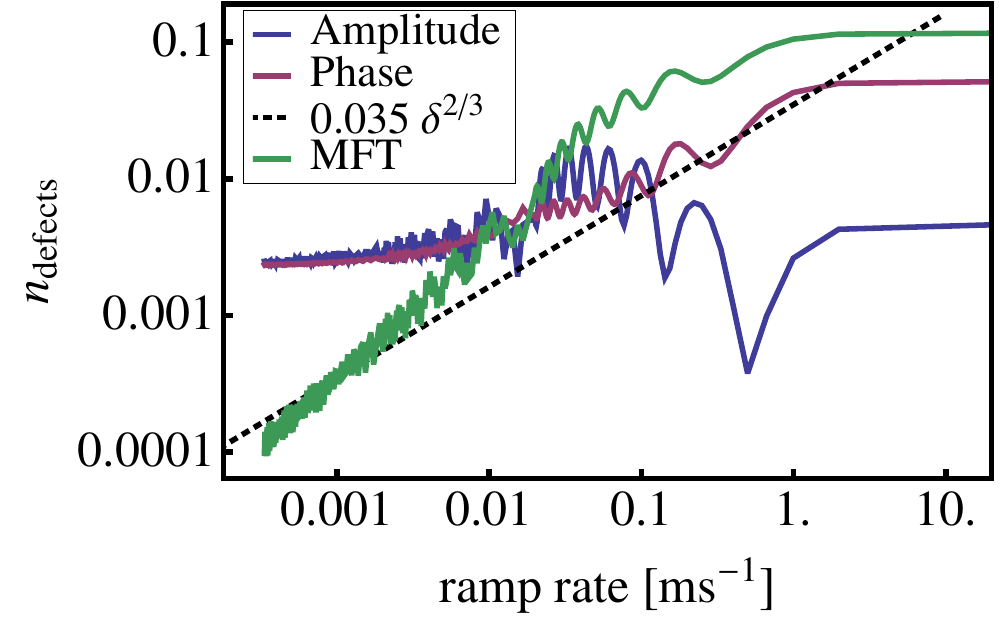}~~
(b)\!\!\!\!\!\!\includegraphics[width=5.5cm]{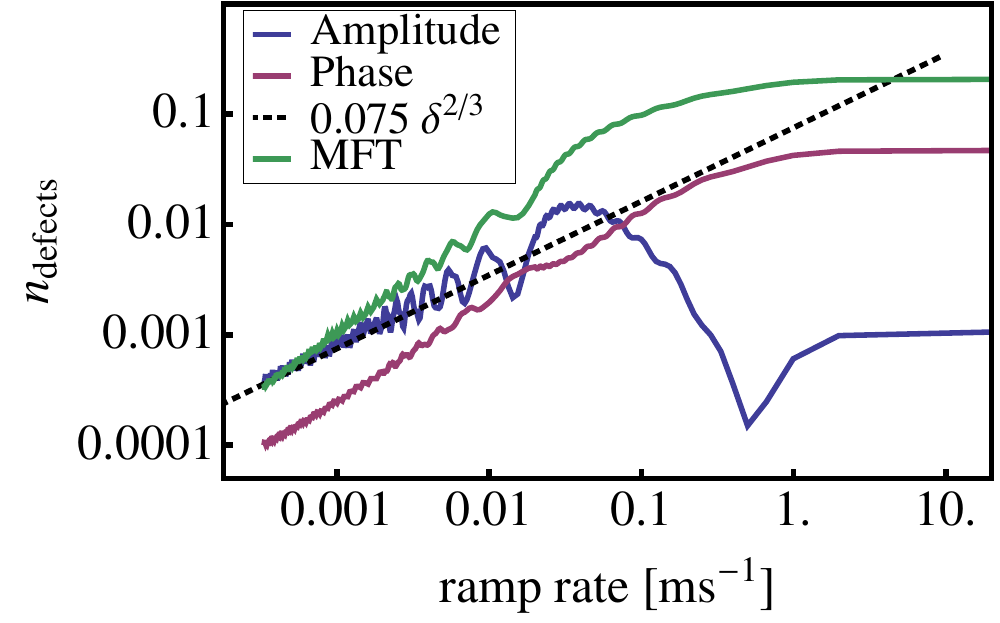}~~
(c)\!\!\!\!\!\!\includegraphics[width=5.5cm]{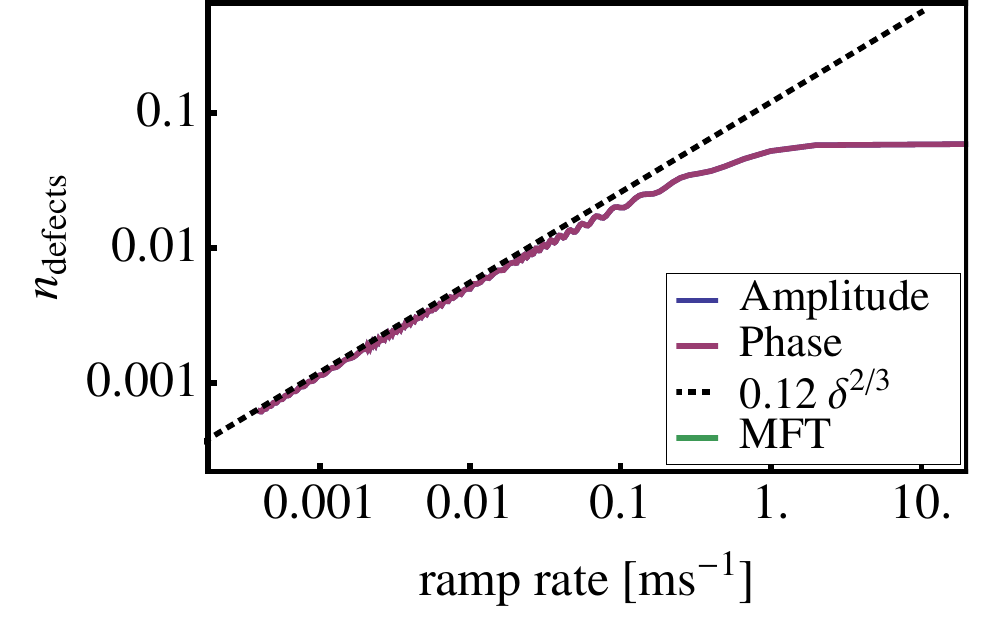}
\caption{Defect density (immediately following the end of the ramp) as a function of the ramp rate ($1/t_\text{ramp}$) on a Log-Log plot. (a) Starting in SF ending in shallow MI. (b) Starting in SF ending in deep MI. (c) Starting at the critical point and ending deep in the MI (here, the MF order parameter is zero throughout). Note that for trace (a), there is a large number of background defects in equilibrium as the ramp only goes into shallow Mott regime, resulting in the truncation of the power law. On the other hand, ending the ramps in the deep Mott regime results in essentially the same trace as the number of excitations.}
\label{fig:d4}
\end{figure*}

To understand the feasibility of observing power-law scaling in experiments, we investigate the defect density. In  Fig.~\ref{fig:d4} we plot the components of the defect density due to the mean-field, the amplitude modes and the phase modes (i.e.~corresponding to the various terms of Eq.~\eqref{eq:Pd}). We begin by looking at SF $\rightarrow$ shallow MI ramps. For these ramps the number of defects created quickly saturates, the saturation values corresponding to the background defect density in the Mott insulator [panel (a) of Fig.~\ref{fig:d4}]. In order to observe scaling within this protocol, it is necessary to subtract of this background defect density. An alternative approach is to perform ramps that go deep into the Mott insulating region and thereby convert excitations to defects. The results of this approach are illustrated in panel (b) of Fig.~\ref{fig:d4}, which shows nice power law scaling in the defect density down to extremely long ramp times ($\sim 3 \, \text{s}$). However, the $2/3$ power law scaling associated with the amplitude mode is partially obscured by the defects associated with the decaying mean field, and only becomes apparent for timescales $1\,\text{s}\sim 15*(2\pi \hbar /J_c)$ and defect densities of 0.1\%. We remark that all issues: (1) background defect density, (2) oscillations, (3) obscuring by defects associated with the mean field and (4) low defect density in the scaling regime can be avoided by starting the ramp at the QCP and going deep into the Mott insulator [panel (c) of Fig.~\ref{fig:d4}]. Using this protocol, the power law scaling sets in at $\sim\frac{1}{6}(2\pi\hbar/J_c)$ and defect density of 1\%.

Due to convenience for the MF+G method, we have attacked the problem of dynamics in the setting of the quantum rotor model. However, we expect that all the qualitative features will transfer directly to the Bose-Hubbard model, as was the case for equilibrium properties. Thus in parametric ramps of the Bose-Hubbard model we expect to find the fast and slow regimes. In the fast regime, we again expect to see response and oscillations on timescale of $2*(2\pi \hbar/U_{H,c})$. While in the slow regime we expect to see a crossover to power law scaling at timescales of $15*(2\pi\hbar/J_{H,c})$ and defect densities of $0.05-0.5\%$.

\section{Discussion}
\label{sec:discussion}
In this Paper, we have investigated the static and dynamic properties of the Bose-Hubbard model, and the related Spin-1 quantum rotor model using four types of methods: (1) Mean Field, (2) Exact Diagonalization, (3) cluster Mean Field, and (4) Mean Field + Gaussian fluctuations. These methods were chosen as they are suitable for studying both statics and dynamics. We show that these methods are reasonably consistent with each other and quantum Monte Carlo calculations where available. The main conclusions are as follows.

For the static case, we can compare these methods to Quantum Monte Carlo simulations. We find that although the qualitative features and trends displayed by these methods are very good, it is difficult to achieve high quantitative accuracy as it requires the use of large system/cluster sizes. In particular, we have applied the methods to the calculation of several observables. These calculations are much simpler than the corresponding quantum Monte Carlo calculations and thus could be of interest to experimentalist for data fitting. We find the following properties. (1) The location of the phase transitions is accessible to the various mean field methods (MF, CMF, and MF+G) but not to the exact diagonalization. The single site mean field makes an errors of 30\%, by going to a $3\times 4$ cluster the error is reduced to 11\%. (2) The defect density in the vicinity of the phase transition varies smoothly within exact diagonalization, but shows a kink within MF and CMF methods. As the defect density is associated with a local observable, we believe that any singular behavior at the phase transition should be strongly muted. Indeed, we see the weakening of the kink within CMF theory as the size of the cluster is increased. Further, the defect density obtained by the CMF method approaches the one obtained by the ED method as the size of the cluster (for CMF) and system (for ED) is increased. (3) Particle-particle correlation functions show signatures of repulsion on nearest-neighbor sites while particle-hole correlation functions show attraction. However, these correlation function strongly depend on the method being used. (4) Finally, at large distances, the particle-particle correlation function shows a diverging correlation length, described quantitatively by Eq.~(\ref{eq:correlation}) in the vicinity of the QCP.

For dynamics, we find two regimes: a fast regime dominated by nearest neighbor physics which is addressed quite well by CMF/ED methods, and a slow regime dominated by long wavelength excitations, which is addressed by MF+G method. The methods we use, unlike QMC, are well suited for treating dynamics. Following the intuition gained by comparing them in equilibrium with each other and with QMC, we suspect that the methods can be used as a guide for understanding experiments. Specifically, for properties like the defect density, we expect the results to be accurate to within a factor of two.

We find that the fast regime is dominated by the peak in the density of states of single quasiparticle modes in the vicinity of $U/2$, which correspond to short-wavelength excitations, and results in timescales of $\sim 2*(2\pi \hbar/U_c)$. As a result, the short time behavior can probably be accurately modeled, even in inhomogeneous systems, using the CMF method. However, one must keep in mind that in equilibrium, the CMF method always underestimates the defect density on the Mott side, and this should be taken into account in interpreting the results of dynamics. 

The slow regime is dominated by long wavelength amplitude modes, which are described by the MF+G method. In this regime we find that universal power law scaling in the number of excitations created as a function of the ramp rate sets in for sufficiently slow ramps. That is the number of excitations (or defects for ramps that go deep into the Mott insulator) goes as $r^{2/3}$ where $r$ is the ramp rate. Quantitatively, the appearance of scaling depends on the protocol being used. Scaling is easiest to see for ramps that start at the QCP and go deep into the Mott regime, as the scaling is not obscured by $2\pi \hbar/J_c$ oscillations nor the decay of the order parameter. Indeed, in this protocol the scaling sets in at $\frac{1}{6} (2 \pi \hbar/J_c)$ timescale. On the other hand, for ramps that  start in SF phase, longer timescales of $\sim 15 *(2 \pi \hbar/J_c)$ are required to observe $2/3$ power law scaling. 

In conclusion, our calculations in equilibrium provide a number of results that can be useful for interpreting experiments, including an intuition for the behavior of defect density near the QCP and a simple expression for the defect-defect correlation function $g_2(l)$ in the vicinity of the QCP. Further, understanding what our methods get right and what they get wrong in equilibrium gives us an understanding of how to apply them to dynamics. Our dynamical calculations provide an understanding of short timescales dominated by the peak in the density of states. Finally, we obtain the defect densities and timescales needed to observe universal power law scaling.

\section{Acknowledgements}
It is our pleasure to thank M. Greiner, W. S. Bakr, and J. Simon for explaining their experimental setup and A. Polkovnikov for his encouragement and valuable insights. We would also like to thank G. Refael for valuable discussions, and the Aspen Center for Physics for it's hospitality. This work was supported by a grant from the Army Research Office with funding from the DARPA OLE program, Harvard-MIT CUA, NSF Grant No. DMR-07-05472, AFOSR Quantum Simulation MURI, AFOSR MURI on Ultracold Molecules, the ARO-MURI on Atomtronics, the Lee A. DuBridge fellowship (DP), DFG grant No 609/1-1 (BW), The Danish National Research Foundation.

\appendix
\begin{widetext}
\section{MF+G in equilibrium}
\label{app:equilibriumMFpG}
The goal of this appendix is to provide details of the modifications to the Mean Field ground state due to Gaussian fluctuations. We begin by giving details of the Bogoliubov transformations, and then construct expressions for the defect density and various correlation functions.

\subsection{Bogoliubov transformations}
\label{app:Bogoliubov}
The quadratic Hamiltonians $H_{\sigma,k}$ are solved by the Bogoliubov transformations
\begin{align}
b^\dag_{\sigma k}&=v_{\sigma k} \gamma_{\sigma -k}+u_{\sigma k} \gamma_{\sigma k}^\dag,
\label{eq:Bogoliubov}
\end{align}
where the coherence factors are
\begin{align}
u_{\sigma k}&=\left(\frac{\alpha_{\sigma,k}}{2\sqrt{\alpha_{\sigma,k}^2-\beta_{\sigma,k}^2}}+\frac{1}{2}\right)^{\frac{1}{2}} &
v_{\sigma k}&=-\text{sgn}(\beta_{\sigma,k})\left(\frac{\alpha_{\sigma,k}}{2\sqrt{\alpha_{\sigma,k}^2-\beta_{\sigma,k}^2}}-\frac{1}{2}\right)^{\frac{1}{2}}
\label{eq:uv}
\end{align}
and the Hamiltonian becomes
\begin{align}
H_{\sigma}&= \frac{1}{2} \sum_k \left(E_{\sigma k}+\frac{1}{2}\right) \gamma_{\sigma k}^\dag \gamma_{\sigma k}&
E_{\sigma k}&=\sqrt{\alpha_{\sigma k}^2-\beta_{\sigma k}^2}. \label{eq:dispersion}
\end{align}
The ground state corresponds to the vacuum of $\gamma_{\sigma,k}$ bosons, or in terms of the $b_{\sigma,k}$ operators to the squeezed state of Eq.~\eqref{eq:squeezedPsi} with 
\begin{align}
c_{\sigma,k}&=v_{\sigma,k}/u_{\sigma,k},& 
\zeta_{\sigma,k}&=0.
\end{align}

\subsection{Particle, hole, and defect density}
\label{app:phd}
To obtain an expression for the particle, hole, and defect density operators in terms of the $b$ operators, we use the same procedure as we used with the Hamiltonian: we first write down the density operator in terms of the $t$ operators of Eq.~\eqref{eq:M}, next we replace the $t$ operators by $b$ operators using the transformation $M$ from Eq.~\eqref{eq:M}, finally we expand the result to second order in $b_{\alpha,i}$ and $b_{\phi,i}$ keeping in mind that we must replace $b_{0,i}^\dagger b_{0,i}=1-b_{\alpha,i}^\dagger b_{\alpha,i} - b_{\phi,i}^\dagger b_{\phi,i}$. Following this procedure we obtain the operators for the particle, hole, and defect density $P_p(i)$, $P_h(i)$, $P_d(i)$
\begin{align}
P_p(i)&=t_{+,i}^\dagger t_{+,i}=\frac{1}{4}\left[1-\cos(\theta)\right]-\frac{1}{4}\sin(\theta)\left[b^\dag_{\alpha i} + b_{\alpha i} \right]+\frac{1}{2} \sin\left(\frac{\theta}{2}\right)\left[b^\dag_{\phi i} +b_{\phi i} \right]\nonumber\\
&+\frac{1}{2}\cos(\theta) b^\dag_{\alpha i}  b_{\alpha i} +\frac{1}{4}\left[1+\cos(\theta)\right] b^\dag_{\phi i}  b_{\phi i} -\frac{1}{2}\cos\left(\frac{\theta}{2}\right) \left[b^\dag_{\alpha i} b_{\phi i} + b_{\phi i}^\dagger b_{\alpha i} \right]\\
P_h(i)&=t_{-,i}^\dagger t_{-,i}=\frac{1}{4}\left[1-\cos(\theta)\right]-\frac{1}{4}\sin(\theta)\left[b^\dag_{\alpha i} + b_{\alpha i} \right]-\frac{1}{2} \sin\left(\frac{\theta}{2}\right)\left[b^\dag_{\phi i} +b_{\phi i} \right]\nonumber\\
&+\frac{1}{2}\cos(\theta) b^\dag_{\alpha i}  b_{\alpha i} +\frac{1}{4}\left[1+\cos(\theta)\right] b^\dag_{\phi i}  b_{\phi i} +\frac{1}{2}\cos\left(\frac{\theta}{2}\right) \left[b^\dag_{\alpha i} b_{\phi i} + b_{\phi i}^\dagger b_{\alpha i} \right]\\\
P_d(i)&=(S^z_i)^2=t_{+,i}^\dagger t_{+,i}+t_{-,i}^\dagger t_{-,i} \nonumber\\
&=\frac{1}{2}\left[1-\cos(\theta)\right]-\frac{1}{2}\sin(\theta)\left[b^\dag_{\alpha i} +b_{\alpha i} \right]+\cos(\theta) b^\dag_{\alpha i}  b_{\alpha i} +\frac{1}{2}\left[1+\cos(\theta)\right] b^\dag_{\phi i}  b_{\phi i} 
\label{eq:Pd}
\end{align}
We note that the expressions we obtain for the density operators satisfy the relation that $P_d(i)=P_p(i)+P_h(i)$. In the following subsection, we shall use the operators $P_p(i)$, $P_h(i)$, $P_d(i)$ to compute their correlation functions. However, in addition to the correlation functions we shall also need to compute the expectation value of the defect density operator $P_d$ in equilibrium in order to compare the result from MF+G method with the other methods. Taking the expectation value of the $P_d(i)$ operator in equilibrium, we obtain
\begin{align}
\langle P_d \rangle =\frac{1}{2}\left[1-\cos(\theta)\right] + \cos(\theta) \sum_k v_{\alpha k}^2 +\frac{1}{2} \left[1+\cos(\theta)\right] \sum_k v_{\phi,k}^2
\end{align}
where the values for the coherence factors in equilibrium are obtained from expressions Eqs.~\eqref{eq:MFgs}, \eqref{eq:alpha:alpha}-\eqref{eq:beta:phi}, \eqref{eq:uv}. To complete the discussion, we give the explicit expression for defect probability in terms of the quadratic Hamiltonian parameters $\alpha_{\sigma,k}$ and $\beta_{\sigma, k}$
\begin{align}
\left\langle P_d(i) \right\rangle &= \sin\left(\frac{\theta}{2}\right) 
+ \left[\cos^2\left(\frac{\theta}{2}\right) -\sin^2\left(\frac{\theta}{2}\right) \right]\int\,\frac{d^2k}{(2\pi)^2} 
\left(\frac{\alpha_{\alpha,k}}{2\sqrt{\alpha_{\alpha,k}^2-\beta_{\alpha,k}^2}}-\frac{1}{2}\right)\nonumber\\
&+ \left[1-\sin^2\left(\frac{\theta}{2}\right) \right]\int\,\frac{d^2k}{(2\pi)^2} 
\left(\frac{\alpha_{\phi,k}}{2\sqrt{\alpha_{\phi,k}^2-\beta_{\phi,k}^2}}-\frac{1}{2}\right).
\end{align}

\subsection{Correlation functions}
\label{app:CF}
In this subsection our goal is to compute the correlation functions
\begin{align}
f_2^\text{p-p}(l)=\langle P_d(i) \rangle^2 (g_2^\text{p-p}(l)-1)=\langle P_p(i) P_p(i+l) \rangle - \langle P_p(i) \rangle \langle P_p(i+l) \rangle,\\
f_2^\text{p-h}(l)=\langle P_d(i) \rangle^2 (g_2^\text{p-h}(l)-1)=\langle P_p(i) P_h(i+l) \rangle - \langle P_p(i) \rangle \langle P_h(i+l) \rangle,\\
f_2^\text{d-d}(l)=\langle P_d(i) \rangle^2 (g_2^\text{d-d}(l)-1)=\langle P_d(i) P_d(i+l) \rangle - \langle P_d(i) \rangle \langle P_d(i+l) \rangle.
\end{align} 
Where, we use $\langle P_d(l)\rangle^2$ to normalize all of the $g_2$'s so that they can be directly compared with each other. In the following, we shall suppress the dependence on position $i$ since we have a homogenous system. 

We begin by obtaining expressions for the $f_2$'s up to fourth order in $b$ operators. Although fourth order may seem to be overkill, we will see that the second order terms disappear on the Mott side due to $\sin(\theta)$ becoming zero thus forcing us to go to the next non-zero order.

On the superfluid side, since $\sin(\theta)\neq 0$ the second order term
\begin{align}
{f^{d-d\, (2)}_2}(l)&=\frac{\sin^2(\theta)}{4}\left\langle \left( b_{\alpha, i}^\dagger + b_{\alpha, i} \right) \left( b_{\alpha, i+l}^\dagger + b_{\alpha, i+l} \right) \right\rangle
\end{align}
is non-zero. Using the zero temperature expressions
\begin{align}
\left\langle b_{\sigma k} b_{\sigma -k} \right\rangle &= \left\langle b_{\sigma k}^\dagger b_{\sigma -k}^\dagger \right\rangle=u_{\sigma k} v_{\sigma k}=-\frac{\beta_{\sigma k}}{2\sqrt{\alpha_{\sigma k}^2-\beta_{\sigma k}^2}},\label{eq:bdbd}\\
\left\langle b_{\sigma k}^\dagger b_{\sigma k} \right\rangle &=v_{\sigma k}^2=\frac{\alpha_{\sigma k}}{2\sqrt{\alpha_{\sigma k}^2-\beta_{\sigma k}^2}}-\frac{1}{2},\\
\left\langle b_{\sigma k} b_{\sigma k}^\dagger \right\rangle &=u_{\sigma k}^2 =\frac{\alpha_{\sigma k}}{2\sqrt{\alpha_{\sigma k}^2-\beta_{\sigma k}^2}}+\frac{1}{2}\label{eq:bbd}.
\end{align}
we obtain
\begin{align}
f_2^{d-d\,(2)}(l)&=\frac{\sin^2(\theta)}{4} \int\,\frac{d^2k}{(2\pi)^2}\cos(l \cdot k)\langle (b_{\alpha, k}^\dag + b_{\alpha, -k})(b_{\alpha, -k}^\dag + b_{\alpha, k})\rangle, \\
&=\frac{\sin^2(\theta)}{4} \int\,\frac{d^2k}{(2\pi)^2} \cos(l \cdot k) \sqrt{\frac{\alpha_{\alpha,k}-\beta_{\alpha,k}}{\alpha_{\alpha,k}+\beta_{\alpha,k}}}. \label{Eq:fdd2}
\end{align}

On the Mott side, $\sin(\theta) = 0$ and the second order term vanishes; therefore, we are forced to look to the fourth order term. Surprisingly, we also find that the contributions from the fourth order term are important on the superfluid side (see main text). We note that the only interesting fourth order term arises as a consequence of the product of the two second order terms in \eqref{eq:Pd} [i.e., $P_d^{(0)}(i) P_d^{(4)}(i+l)$ is independent of $l$ and is therefore canceled by the regular part;  $P_d^{(1)}(i) P_d^{(3)}(i+l)$ vanishes due to $\sin(\theta) = 0$]. Applying Wick's theorem, which we are allowed to do since we are working within the quadratic approximation to the Hamiltonian, we obtain
\begin{align}
&f_2^{d-d\,(4)}(l)= \cos^2(\theta) \left[ \left\langle b_{\alpha, i}^\dagger b_{\alpha, i+l}^\dagger \right\rangle \left\langle b_{\alpha, i} b_{\alpha, i+l} \right\rangle + \left\langle b_{\alpha, i}^\dagger b_{\alpha, i+l} \right\rangle \left\langle b_{\alpha, i}^\dagger b_{\alpha, i+l} \right\rangle \right]\nonumber\\
&\quad\quad +\left(\frac{1+\cos(\theta)}{2}\right)^2 \left[ \left\langle b_{\phi, i}^\dagger b_{\phi, i+l}^\dagger \right\rangle \left\langle b_{\phi, i} b_{\phi, i+l} \right\rangle + \left\langle b_{\phi, i}^\dagger b_{\phi, i+l} \right\rangle \left\langle b_{\phi, i}^\dagger b_{\phi, i+l} \right\rangle \right]\\
&=\cos^2(\theta) \left[F_\alpha^2+G_\alpha^2\right]+\frac{1}{4} \left(1+\cos(\theta)\right)^2\left[F_\phi^2+G_\phi^2\right]\label{Eq:f24}
\end{align}
where we have introduced the notation
\begin{align}
F_\sigma&=\int\,\frac{d^2k}{(2\pi)^2}\frac{\beta_{\sigma,k} \cos(l\cdot k) }{2\sqrt{\alpha_{\sigma,k}^2-\beta_{\sigma,k}^2}} \label{Eq:Fs} & 
G_\sigma&=\int\,\frac{d^2k}{(2\pi)^2}\frac{\alpha_{\sigma,k}\cos(l\cdot k) }{2\sqrt{\alpha_{\sigma,k}^2-\beta_{\sigma,k}^2}}
\end{align}
By going to fourth order in fluctuations for correlations but only to second order in the Hamiltonian we miss another term that is fourth order in fluctuations, namely fourth order on site $i$ and zeroth order on site $j$ and vice versa. As we already explained, this term does not contribute to $f_2$ and only enters $g_2$ via a slight modification of the defect probability itself.

We can find similar expressions for the particle-particle, particle-hole, and hole-hole correlation functions:
\begin{align}
f_2^{p-p\, (2)}(l)=f_2^{h-h(4)}(l)&=\frac{1}{8}\sin^2(\theta)\left[F_\alpha+G_\alpha\right]+\frac{1}{2}\sin^2\left(\frac{\theta}{2}\right)\left[F_\phi+G_\phi\right],\\
f_2^{p-h\, (2)}(l)=f_2^{h-h(4)}(l)&=\frac{1}{8}\sin^2(\theta)\left[F_\alpha+G_\alpha\right]-\frac{1}{2}\sin^2\left(\frac{\theta}{2}\right)\left[F_\phi+G_\phi\right],
\end{align}
and
\begin{align}
f_2^{p-p\, (4)}(l)=f_2^{h-h(4)}(l)&=\frac{1}{4}\cos^2(\theta)\left[F_\alpha^2+G_\alpha^2\right]
+\frac{1}{16} \left(1+\cos(\theta)\right)^2\left[F_\phi^2+G_\phi^2\right]+\frac{1}{2}\cos^2\left(\frac{\theta}{2}\right)\left[F_\alpha F_\phi+G_\alpha G_\phi\right],\\
f_2^{p-h\, (4)}(l)=f_2^{h-h(4)}(l)&=\frac{1}{4}\cos^2(\theta)\left[F_\alpha^2+G_\alpha^2\right]
+\frac{1}{16} \left(1+\cos(\theta)\right)^2\left[F_\phi^2+G_\phi^2\right]-\frac{1}{2}\cos^2\left(\frac{\theta}{2}\right)\left[F_\alpha F_\phi+G_\alpha G_\phi\right].
\end{align}

\section{Evolution via quadratures}
\label{app:quads}
An alternative view to writing down the Schrodinger equation for the wave function, Eq.~\eqref{eq:Schrodinger}, is to study directly the evolution of operators of interest using the Heisenberg equations of motion. Since the Hamiltonian is always quadratic, the Heisenberg equations of motion for the quadratures $\langle b^\dagger_{\sigma, k} b_{\sigma, k} + b^\dagger_{\sigma, -k} b_{\sigma, -k} \rangle$, $\langle b^\dagger_{\sigma, k} b^\dagger_{\sigma, -k} \rangle$, and $\langle b_{\sigma, k} b_{\sigma, k} \rangle$ close on themselves. Taking into account the Berry phase, the Heisenberg equations of motion are
\begin{align}
\partial_t \langle b^\dagger_{\sigma, k} b_{\sigma, k} + b^\dagger_{\sigma, -k} b_{\sigma, -k} \rangle &= \frac{i}{\hbar} \left\langle \left[b^\dagger_{\sigma, k} b_{\sigma, k} + b^\dagger_{\sigma, -k} b_{\sigma, -k}, H_{\sigma,k} \right] \right\rangle \nonumber\\
&=\frac{2 i}{\hbar} \left( \beta_{\sigma,k}(t) \langle b^\dagger_{\sigma, k} b^\dagger_{\sigma, -k} \rangle - \beta^*_{\sigma,k}(t)  \langle b_{\sigma, k} b_{\sigma, -k} \rangle \right),\\
\partial_t \langle b^\dagger_{\sigma, k} b^\dagger_{\sigma, -k} \rangle &= \frac{i}{\hbar} \left(-2 \alpha_{\sigma,k}(t) \langle b^\dagger_{\sigma, k} b^\dagger_{\sigma, -k} \rangle - \beta^*_{\sigma,k}(t) \langle b^\dagger_{\sigma, k} b_{\sigma, k} + b^\dagger_{\sigma, -k} b_{\sigma, -k} \rangle\right) \nonumber \\
&\quad\quad\quad\quad\quad\quad\quad\quad\quad\quad\quad\quad\quad\quad\quad\quad+ 2 A^\dagger_{\sigma,k}(t) \langle b^\dagger_{\sigma, k} b^\dagger_{\sigma, -k} \rangle,\\
\partial_t \langle b_{\sigma, k} b_{\sigma, -k} \rangle &= \frac{i}{\hbar} \left(2 \alpha_{\sigma,k}(t) \langle b_{\sigma, k} b_{\sigma, -k} \rangle + \beta_{\sigma,k}(t) \langle b^\dagger_{\sigma, k} b_{\sigma, k} + b^\dagger_{\sigma, -k} b_{\sigma, -k} \rangle\right) \nonumber \\
&\quad\quad\quad\quad\quad\quad\quad\quad\quad\quad\quad\quad\quad\quad\quad\quad+ 2 A_{\sigma,k}(t) \langle b_{\sigma, k} b_{\sigma, -k} \rangle.
\end{align}
The initial conditions for the quadratures at zero temperature can be obtained from Eqs.~\eqref{eq:bdbd}-\eqref{eq:bbd}. A useful remark is that due to the quadratic nature of the effective Hamiltonian, the thermal as well as ground states of the system can be expressed by specifying just the three quadratures appearing in the above equations. Moreover, the equations of motion for the quadratures still hold for the thermal state. However, to take finite temperature into account we must modify the MF solution on top of which we develop the Gaussian fluctuations, i.e. $\theta(t)$ and $\phi(t)$, as well as the initial conditions.

\section{Normalized correlation functions and correlation functions in dynamics}
\label{app:CFN}

\begin{figure*}
\begin{minipage}[b]{0.2cm}
	{\bf (a)}
	
	\vspace{2.8cm}
\end{minipage}
\begin{minipage}[t]{5.6cm}
	\includegraphics[scale=0.55]{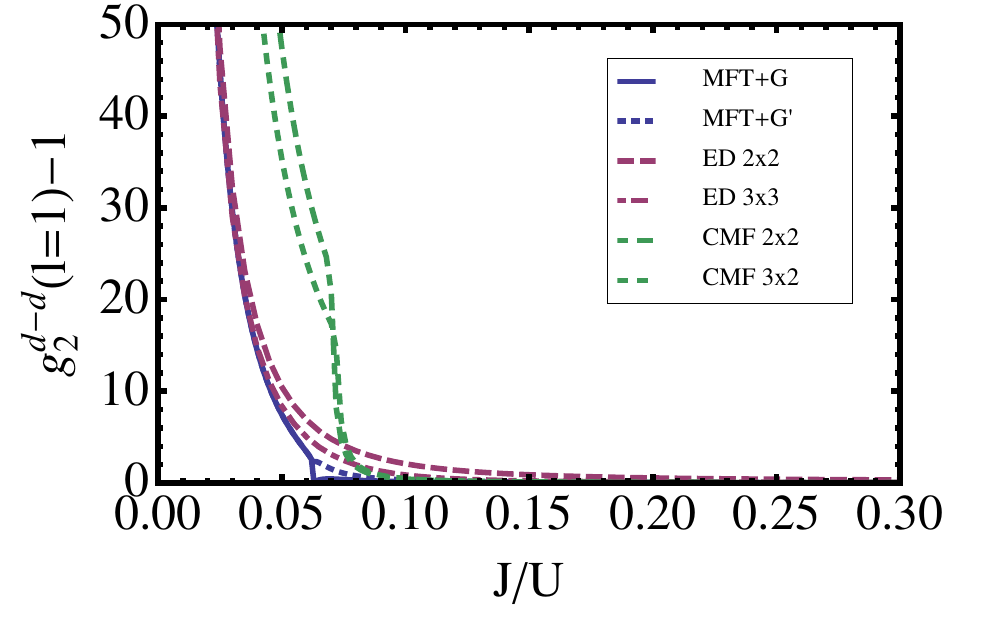}
\end{minipage}
\begin{minipage}[b]{0.2cm}
	{\bf (b)}
	
	\vspace{2.8cm}
\end{minipage}
\begin{minipage}[t]{5.6cm}
	\includegraphics[scale=0.55]{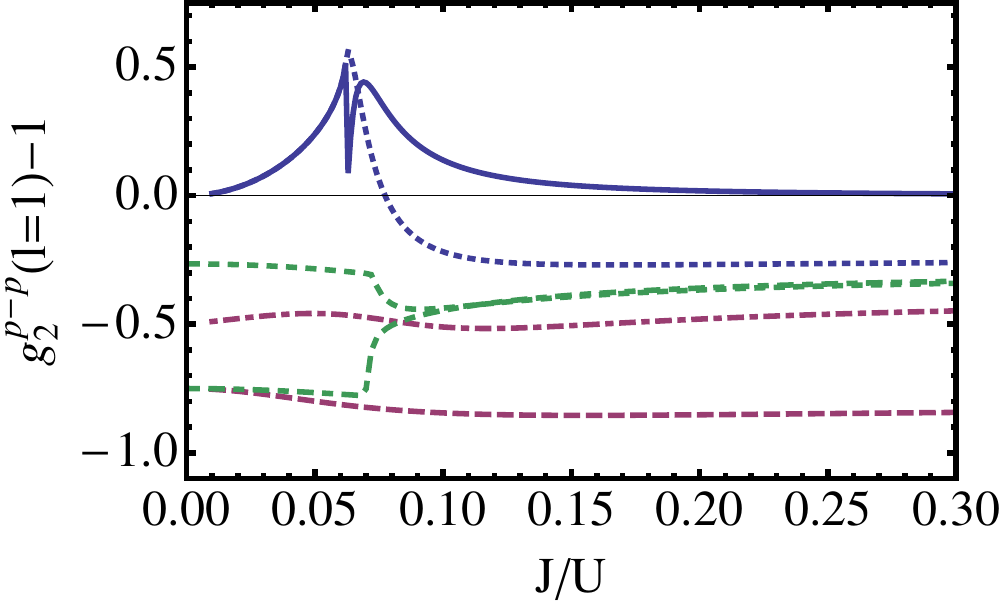}
\end{minipage}
\begin{minipage}[b]{0.2cm}
	{\bf (c)}
	
	\vspace{2.8cm}
\end{minipage}
\begin{minipage}[t]{5.6cm}
	\includegraphics[scale=0.55]{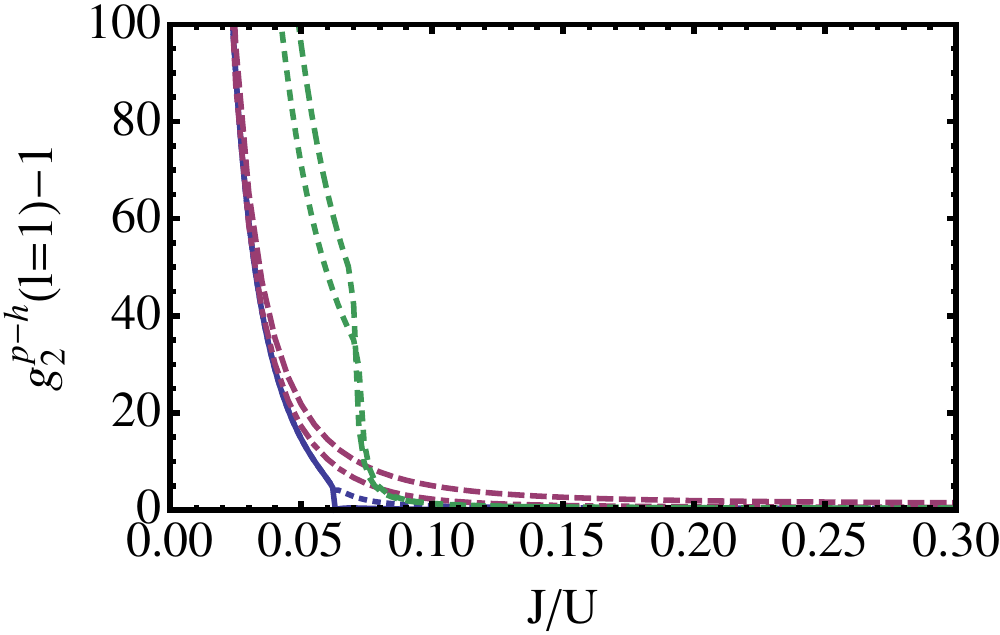}
\end{minipage}
\caption{Analogous to Figure~\ref{Fig:f2l1} in the main text, but with normalized correlation functions:\\ 
(a) $g_2^{d-d}(l=1)-1\equiv  \frac{\langle P_d(l) P_d(0)\rangle-\langle P_d(0) \rangle^2}{\langle P_d(0) \rangle^2}$, \\
(b) $g_2^{p-p}(l=1)-1\equiv  \frac{\langle P_p(l) P_p(0)\rangle-\langle P_p(0) \rangle^2}{\langle P_p(0) \rangle^2}$, \\
(c)  $g_2^{p-h}(l=1)-1\equiv  \frac{\langle P_p(l) P_h(0)\rangle-\langle P_p(0) \rangle \langle P_h(0) \rangle}{\langle P_p(0) \rangle \langle P_h(0) \rangle}$.}
\label{fig:g2l1aa}
\end{figure*}

\begin{figure*}
\includegraphics[width=15cm]{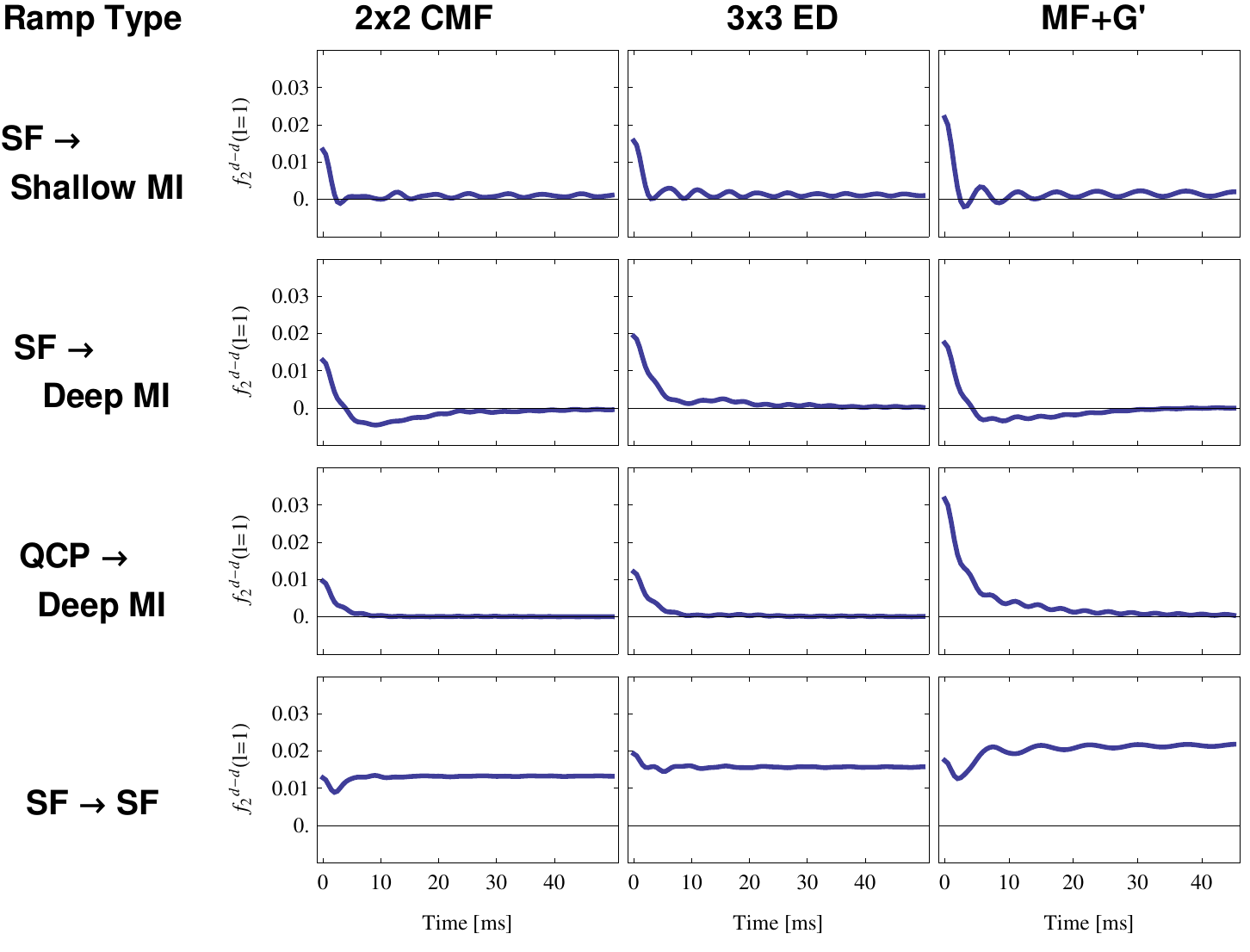}
\caption{Nearest neighbor correlation function $f^{d-d}_2(l=1)$ (immediately following the end of the ramp) as a function of the ramp time. The calculations were performed for four different ramp types (as indicated by row headings), using the three (applicable) methods (as indicated by column headings). Parameters used in calculations are identical to those of Fig.~\ref{fig:d1}. }
\label{fig:f2dynamics} 
\end{figure*}

In this appendix we supplement the main text by (1) plotting the normalized equilibrium correlation functions, e.g. $g^{d-d}_2=f^{d-d}_2/\langle P_d(i)\rangle^2+1=\langle P_d(i) P_d(i+1) \rangle/\langle P_d(i)\rangle^2$ [Fig.~\ref{fig:g2l1aa}] and (2) plotting the correlation function $f^{d-d}_2(l=1)$ (obtained using ED, CMF, and MF+G' methods) at the end of the ramp [Fig.~\ref{fig:f2dynamics}].

\section{Connection with Quantum Gas microscope experiments}
\label{app:exp}
The absolute timescales involved in experiments depend on the details of the implementation. However, the use of $^{87}$Rb atoms is popular, and thus many setups will have similar timescales. We have used the setup of Ref~\onlinecite{Bakr2010} throughout the manuscript to compute typical timescales in terms of seconds. The details, which we now provide, that are necessary to make the connection are the dependence of the hopping matrix element $J$ and the onsite repulsion $U$ on the optic lattice depth $V$. Explicitly, we find
\begin{align}
J(V)&=244.5 \exp(-0.209 V)\, \text{Hz},\\
U(V)&=(83.4 + 23.1V - 0.2985 V^2)\, \text{Hz},
\end{align}
where $V$ is measured in units of recoil energy.
\end{widetext}

\bibliography{BoseHubbardStrip}

\end{document}